\documentclass[epj]{svjour}

\usepackage{graphics}
\usepackage{amssymb,amsmath}
\newcommand{\be}{\begin{equation}}
\newcommand{\ee}{\end{equation}}
\newcommand{\bea}{\begin{eqnarray}}
\newcommand{\eea}{\end{eqnarray}}

\newcommand{\vp}{{\mathbf{p}}}

\newcommand{\ba}{\begin{array}}
\newcommand{\ea}{\end{array}}

\newcommand{\lb}{\label}

\newcommand{\re}[1]{(\ref{#1})}
\newcommand{\var}{\varepsilon}
\newcommand{\varp}{\varepsilon_{\perp}}
\newcommand{\vara}{\varepsilon_{\along}}
\def\trans{\mbox{\tiny$\bot$}} % Transverse component
\def\along{\mbox{\tiny$\|$}}   % Longitudinal component

\newcommand{\ve}[1]{\mathbf{#1}}
\newcommand{\ds}{\displaystyle}

\begin{document}
\title{Dynamical Schwinger effect and high-intensity lasers}
\subtitle{Realising nonperturbative QED}
\author{D.B. Blaschke\inst{1,2,3} \and A.V. Prozorkevich\inst{4} \and
G. R\"opke\inst{3} \and C.D.~Roberts\inst{5} \and
S.M. Schmidt\inst{6}
\and D.S. Shkirmanov\inst{4}
\and S.A.~Smolyansky\inst{4}
% \thanks is optional - remove next line if not needed
%\thanks{\emph{Present address:} Insert the address here if needed}%
}                     % Do not remove
\offprints{David Blaschke} %blaschke@ift.uni.wroc.pl}
% Insert a name or remove this line
%
\institute{Institute for Theoretical Physics, University of Wroc{\l}aw,
Max Born pl. 9, 50-204 Wroc{\l}aw, Poland %
\and Bogoliubov Laboratory for Theoretical Physics, Joint Institute for
Nuclear Research, RU-141980,
Dubna, Russia %
\and Institut f\"ur Physik, Universit\"at Rostock, D-18051 Rostock, Germany
\and Saratov State University, RU-410026, Saratov, Russia %
\and Physics Division, Argonne National Laboratory, Argonne,
IL 60439-4843, USA
\and Forschungszentrum J\"ulich GmbH, D-52428 J\"ulich, Germany
}
\date{Received: November 17, 2008 / Revised version: date}
% The correct dates will be entered by Springer
%
\abstract{We consider the possibility of experimental verification of vacuum $e^+e^-$ pair creation at the focus of two counter-propagating optical laser beams with intensities 10$^{20}$-10$^{22}$ W/cm$^2$, achievable with present-day petawatt lasers, and approaching the Schwinger limit: 10$^{29}$ W/cm$^2$ to be reached at ELI.
Our approach is based on the collisionless kinetic equation for the evolution of the $e^+$ and $e^-$ distribution functions governed by a non-Markovian source term for pair production.
As possible experimental signals of vacuum pair production we consider $e^+e^-$ annihilation into $\gamma$-pairs and the refraction of a high-frequency probe laser beam by the produced $e^+e^-$ plasma.
We discuss the dependence of the dynamical pair production process on laser wavelength, with special emphasis on applications in the X-ray domain (X-FEL), as well as the prospects for $\mu^+\mu^-$ and $\pi^+\pi^-$ pair creation at high-intensity lasers.
We investigate perspectives for using high-intensity lasers as ``boosters'' of ion beams in the few-GeV per nucleon range, which is relevant, e.g., to the exploration of the QCD phase transition in laboratory experiments.
 % end of PACS codes
} %end of abstract
\maketitle
\section{Introduction \label{intro}}
Vacuum $e^+ e^-$ pair creation by a classical electric field is a longstanding prediction in QED \cite{Sauter,HE,JS}. 
A complete theoretical description of the effect exists \cite{Grib,Nick,Greiner,Fradkin}, but there is still no experimental verification. 
The main obstacle is the high value of the critical electric field strength for pair creation; viz., $E_{\rm cr} = m^2/e =  1.3\times 10^{16}$ V/cm for electron-positron
case\footnote{We use $\hbar=c=k_B=1$ throughout.}.
According to the so-called Schwinger formula \cite{JS}, the pair creation rate in a constant electric field,
\begin{equation}
\label{Schwinger}
{S}^{{\,\rm cl}}=\frac{e^2E^2}{4\pi^3}\exp \bigg(-\frac{\pi m^2}{|eE|}\bigg)~,
\end{equation}
is suppressed exponentially when $E\ll E_{\rm cr}$, see Appendix \ref{app:KE}.  However, a very different situation occurs when the field acts only in a finite time interval (dynamical Schwinger effect) \cite{Grib,Smol,Rob,Pop}.  In this case, the Schwinger
formula, as well as its analog for a monochromatic field (Brezin-Itzykson formula \cite{Brezin70}), become inapplicable in the weak field regime.

A few examples have been discussed of physical situations where the Schwinger effect could occur despite the high critical field strength; e.g., relativistic heavy ion collisions \cite{Casher}, neutron stars \cite{magn,Ruffini} and focused laser pulses \cite{focus}.

Since the Schwinger effect is non-perturbative and it requires an exact solution of the dynamical equations it is customary to approximate the complicated structure of a real laser field by a spatially uniform time-dependent electric field.  According to
different estimates \cite{Pop,Brezin70,Casher,magn,Ruffini,focus,Troup,Marinov,Bunkin,BulanovSS}
the effect of vacuum pair creation is unlikely to be observable with presently available laser parameters.

However, recent developments in laser technology, in particular the invention of the chirped pulse amplification method, have resulted in a huge increase in the light intensity at the laser focal spot \cite{Mou,BulanovSV}.  On this basis the European Extreme Light Infrastructure (ELI) project \cite{ELI} will be developed in order to
provide radiation beams of femto- to atto-second duration in the deeply relativistic regime, exceeding intensities of 10$^{25}$ W/cm$^2$.  On the other hand, construction of X-ray free electron lasers {XFEL} \cite{Ring,XFEL} based on the SASE principle is underway at DESY Hamburg.  Thus an experimental verification of the Schwinger effect is coming within reach.

Under conditions of short duration pulses time-dependent effects become important. 
Therefore in our work \cite{PRL06} we use a kinetic equation approach, which allows us to consider the dynamics of the vacuum pair creation process while accounting properly for the initial conditions \cite{Smol,OR}. 
Compared to alternative treatments, this approach is essentially nonperturbative and contains new dynamical aspects, such as longitudinal momentum dependence in the
distribution functions and non-Markovian character of the time evolution. 
It also takes into account the effects of field switching and particle statistics \cite{our,Bloch:1999eu,quark}. 
This approach has been applied already to the periodical field case \cite{Rob} with near-critical values of the field strength and X-ray frequencies. 
In particular, it was shown that there is an accumulation effect when the intensity of the field is about half critical: the average density of pairs grows steadily with
increasing number of field periods. 
The method \cite{Smol} also found application in describing the pre-equilibrium evolution of a quark-gluon plasma produced in ultrarelativistic heavy ion collisions at RHIC and LHC \cite{our,Bloch:1999eu,quark}.  
A characteristic feature of the kinetic approach is the possibility of a description of quasiparticle excitations during all stages of the external field evolution.

\section{Time-dependent Schwinger mechanism \label{sec:2}}
\subsection{Kinetic approach \label{sec:2:1}}
The basic quantity addressed in the kinetic approach \cite{Smol} is the distribution function of quasiparticles in the momentum representation $f(\mathbf{p},t)$
\cite{Grib,Nick,Greiner,Fradkin}.  The kinetic equation for this function can be derived from the Dirac or Klein-Gordon equations in an external time-dependent electric field by the canonical Bogoliubov transformation method \cite{Grib} or in the oscillator representation approach \cite{OR}.  This procedure is exact but it is valid only for the simplest field configurations; e.g., for a spatially-uniform time-dependent electric field with fixed direction $\mathbf{E}(t)=(0,0,E(t))$.  It is assumed that the electric field vanishes at the initial time $t = t_0$, where real particles are absent (``in-vacuum'' state). The derivation of the corresponding kinetic equation (in the collisionless limit) was presented in  Ref.\,\cite{Smol} and is summarized in the Appendix \ref{app:KE}.  The result is
\begin{multline}\label{ke}
\frac{\partial f(\mathbf{p},t)}{\partial t} + e
\mathbf{E}(t)\frac{\partial f(\mathbf{p},t)}{\partial
\mathbf{p}} = S(\mathbf{p},t)~,\\
S(\mathbf{p},t) = \frac12 \Delta(\mathbf{p},t,t)\int\limits_{t_0}^t \! dt_1 \,
\Delta(\mathbf{p},t_1,t)\left[ 1 \pm 2 f(\mathbf{p},t_1)\right] \\
\times\cos{\left[
2\int\limits_{t_1}^{t}dt_2\,\varepsilon(\mathbf{p},t_2,t)\right]},
\end{multline}
where $\mathbf{p}$ is the kinematic momentum and %$\mathbf{p}(t_1,t_2)$ is field-dependent
\begin{align}
\mathbf{p}(t_1,t_2) &= \mathbf{p} - e\int\limits_{t_1}^{t_2}
\mathbf{E} (t')dt',
\nonumber\\
\Delta(\mathbf{p},t_1,t_2)& = e E(t_1)\frac{\varepsilon_{\trans}}
{\varepsilon^2(\mathbf{p},t_1,t_2)}, \ \ s=\frac{1}{2}, \nonumber \\[3pt]
\Delta(\mathbf{p},t_1,t_2)& = e E(t_1)\frac{p_{\mbox{\tiny$\parallel$}}(t_1,t_2)}
{\varepsilon^2(\mathbf{p},t_1,t_2)}, \ \ s=0, \nonumber \\[3pt]
\varepsilon_{\trans} &= \sqrt{m^2+p_{\mbox{\tiny$\bot$}}^2},  \nonumber \\[3pt]
\varepsilon(\mathbf{p},t_1,t_2)&
=\sqrt{\varepsilon_{\trans}^2 +
p^2_{\mbox{\tiny$\parallel$}}(t_1,t_2)},
\label{kinema}
\end{align}
wherein $m$ and $e$ are the particle's mass and charge, and $s$ is the particle's spin.  The created particles are accelerated by $E_{\rm ext}(t)$, and the associated currents generate an opposing electric field, $E_{\rm int}(t)$.  The total field $E(t)$, which is finally responsible for particle production, is defined as the sum of the external (laser) field $E_{\rm ext}$ and the self-consistent internal field $E_{\rm int}$, which can be found from the Maxwell equation
\begin{multline}
\label{max}
\dot E_{\rm int}(t)= - \frac{e\, 2^{2s}}{(2\pi)^3}\int \frac{d
\mathbf{p}}{\varepsilon (\ve{p})}\left\{
2\,p_{{\scriptscriptstyle\parallel}}\, f(\mathbf{p},t)   \right.\\ +
(\varepsilon_{\trans})^{2s} \int\limits_{t_0}^t \! dt_1 \,
\Delta(\mathbf{p},t_1,t)\left[ 1 \pm 2 f(\mathbf{p},t_1)\right] \\
\left.  \times\cos{\left[
2\int\limits_{t_1}^{t}dt_2\,\varepsilon(\mathbf{p},t_2,t)\right]}
\right\}\ ,
\end{multline}
where $\varepsilon(\ve{p}) = \varepsilon(\mathbf{p},t,t) =
\sqrt{m^2+{\mathbf{p}}^2}$.

In general the kinetic equation involves in addition to the source term, $S(\mathbf{p},t)$, its coupling to Maxwell's equation, which provides for a field-current feedback typical of plasmas, and collision terms.  The importance of these terms depends on the magnitude of the background field and the mass of the produced particles \cite{bastirev}.  For relatively weak fields we expect the produced-particle number density to be small and hence collisions to be rare.  We therefore neglect the collision
term in this work.
Quantum statistics affect the production rate through the term $[1 \pm 2 \,f]$ in Eq.~(\ref{ke}), which ensures that no momentum state has more than one spin-up and one spin-down fermion.  In addition, both this factor and the ``$\cos$'' term introduce non-Markovian character to the system: the first couples in the history of the distribution function's time evolution; the second, that of the field.

% 3 ODE
Equation~\re{ke} can be transformed to a system of three ordinary differential equations for the functions $f$, $u$ and $w$  \cite{Grib},
%written in the canonical momentum variable $\mathbf{q}$
\begin{eqnarray}\label{ode}
 \frac{\partial f}{\partial t} &=& \frac12~\Delta u, \nonumber\\
 \frac{\partial u}{\partial t} &=& \Delta~(1\pm 2f)-2\varepsilon~w,\nonumber\\ 
 \frac{\partial w}{\partial t} &=& 2\varepsilon~u ~,
\end{eqnarray}
where we have suppressed the arguments.  In Eq.\,(\ref{ode}), $\varepsilon = \sqrt{\varp^2 + [ q_{\along} - eA(t) ]^2} $, $ E(t) = \dot{A}(t)$ and the source function $\Delta$ is defined in Eq.\,(\ref{kinema}). 
Another form of this set of equations, which includes the Vlasov dynamics, is met, e.g., in the treatment of the refraction of a probe laser, see Eqs.\,(\ref{refr:11}) of Subsect.\,\ref{sect:3:refr}.
%Nevertheless, its direct integration  is extremely complicated.
The presence of two incommensurable time scales complicates the problem of direct integration of the system: the characteristic time of pair creation is $1/m$ and that of field change, $1/\nu$, where $\nu$ is the laser frequency. Furthermore, in our case Eq.\,\re{ke} contains two small parameters: $E\ll m^2$ and $\nu\ll m$, but we cannot
construct any perturbation theory because of the memory effects present in the argument of the cosine.  These memory effects can only be neglected when
\begin{equation}\label{mem_cond}
    \frac{eE_m}{m \nu} \ll \frac{\nu}{m}
\end{equation}
but this condition would contradict the quasiclassical condition for the electric field; i.e., $E\gg \nu^2$.

Maxwell's equation, (\ref{max}), can be written in the manner of Eq.\,\re{ode}:
\begin{eqnarray}
   \dot E_{\rm in} &=& - j, \nonumber \\
   j &=&  \frac{e}{2 \pi^3}\int \frac{d\mathbf{p}}{\varepsilon}\biggl[ p_{\along} f + \frac12 \varp u \biggr], \ s=\frac{1}{2} ,\nonumber \\
      j &=&  \frac{e}{4 \pi^3}\int\ d\mathbf{p} \frac{p_{\along}}{\varepsilon}\biggl[f + \frac12 u \biggr], \ \ s=0. \label{current}
  \end{eqnarray}
In our case, $E\ll E_{\rm cr}$, the feedback field, $E_{\rm int}$, is negligible and $f \ll 1$ with high accuracy.  Under these conditions the analytic solution of the kinetic equation is
\begin{multline}
\label{fact}
f(\mathbf{p},t) \\ = \left| \,\frac12\, \int\limits_{t_0}^t dt_1 \,
\Delta(\mathbf{p},t_1,t)
\exp{\left(2i\int\limits^{t_1}_{t_0} dt_2 \varepsilon
(\mathbf{p},t_2,t)\right)}\right|^{\,2} \,.
\end{multline}

The total quasiparticle number and energy density are defined as a moments of the distribution function:
\begin{eqnarray}
\label{dens}
n(t)&=& \frac{1}{4 \pi^3} \int d \mathbf{p} f(\mathbf{p},t)\ , \\
\epsilon (t)&=& \frac{1}{4 \pi^3} \int d \mathbf{p}\ \var(\mathbf{p},t)\
f(\mathbf{p},t)\ .
\label{epsil}
\end{eqnarray}
It is well known \cite{Grib} that some of these quantities are divergent and need regularization, which is discussed in Subsects.~\ref{sect:2:r}
and \ref{sect:3:refr}.

\subsection{Asymptotic expansions and regularization \label{sect:2:r}}
In order to regularize divergent expressions for moments of the distribution
functions, such as (\ref{epsil}), we suggest to apply the following procedure \cite{Mamaev78,Zeld71}.  Let us expand the solutions of Eqs.\,\re{ode} as an asymptotic series in $1/\var(\ve{q})$, where $\var(\ve{q})$ is given in Eq.\,(\ref{kinema}):
\begin{align}\label{1:3:series}
  f \sim\sum\limits_n f_n,\quad u\sim\sum\limits_n u_n \quad w\sim\sum\limits_n w_n, \nonumber \\
  f_n\sim u_n \sim w_n \sim \var^{-n},\quad |q|\to \infty,
\end{align}
so that for any $N\ge 1$
\begin{equation}\label{1:3:as_ser}
f-\sum\limits_{n=1}^{N} f_n = o(f_N) \quad (\var \to\infty).
\end{equation}

Substituting these expansions in Eqs.\,\re{ode} and separating the corresponding orders we obtain for the leading terms
\begin{equation}\label{leader}
w_2=\frac{\Delta}{2\var}\ , \qquad
u_3=\frac{\dot{\Delta}}{4\var^2} , \qquad f_4 =
\left(\frac{\Delta}{4\var}\right)^2.
\end{equation}
Hence the number density integral is convergent but the current and energy
densities are logarithmically divergent.  The regularization procedure suggested in \cite{Mamaev78,Zeld71} consists in the substraction of the leading terms, Eq.\,\re{leader}, from the integrand:
\begin{eqnarray}\label{1:3:renorm}
j_R(t)&=& \frac{e 2^{2s}}{4 \pi^3}
\int d\mathbf{p} \frac{p_{\along}}{\var}\,\left[f + \frac12 (u-u_3)
\left(\frac{\varp}{p_{\along}}\right)^{2s}\right], \nonumber\\
\varepsilon_R (t) &=& \frac{1}{4 \pi^3} \int d \mathbf{p}\, \var (f - f_4).
\end{eqnarray}
These expressions obey the energy conservation law without the external field.

The regularization presented in Eqs.\,\re{1:3:renorm} is equivalent to charge regularization in QED \cite{Mamaev78,Zeld71}.  Let us write the counter-term for the current density in the form
\begin{equation}\label{1:3:6}
e^2\, \dot E(t) \int\limits_{|p|>\Lambda} d\mathbf{p}
\frac{p_{\along}^2}{4\var^5}
\left(\frac{\varp}{p_{\along}}\right)^{2s} \equiv \dot E(t) U_3(t),
\end{equation}
where $U_3(t)$ is a divergent integral and $\Lambda$ is some scaling factor
with dimension energy.  We add the expression in Eq.\,\re{1:3:6} to both sides of the Maxwell equation and denote the regularization constant as $Z=(1+U_3)^{-1}$, then
\begin{equation}\label{1:3:6a}
  Z^{-1} \dot{E} = -j + V_3 \dot{E}.
\end{equation}
Using the standard definitions of renormalized quantities according to the Ward identity: $eA=e_R A_R$,
\begin{equation}\label{1:3:7}
  e_R=Z^{1/2} e, \qquad E_R= Z^{-1/2}E ,
\end{equation}
we obtain the renormalized Maxwell equation
\begin{equation}\label{1:3:8}
  \dot{E}_R=-j_R + \dot{E}_R V_{3R}.
\end{equation}

Formally the regularization in Eq.\,\re{1:3:renorm} is ambiguous because the parameter $\Lambda$ is arbitrary.  The choice $\Lambda=0$, which was used in \cite{Mamaev78,Zeld71}, is not a preferred value of any reason.
In our problem we have other energy-scales, e.g. ($m, eA$), which have a definite physical meaning.  Choosing $\Lambda \gg m, \ \Lambda\gg eA$, is sufficient for the current regularization
\begin{equation}\label{1:3:9}
U_3=\frac{(eg_s)^2}{6\pi^2} \int\limits_{\Lambda}^{\infty} \frac{dx}{x}\,.
\end{equation}
The inclusion of counter-terms in the region $p\to 0$ can eventually result in
some non-physical consequences as, e.g., in a negative energy density from 
Eq.\,\re{1:3:renorm}.

\section{Effects to be tested in experiment \label{sec:3}}
\subsection{Photon production  \label{sect:3:1}} %\cite{PRL06}}
In this section we consider the region of field parameters already achievable at currently operating laser systems \cite{Jena,SLAC}; namely, $\nu^2 \ll E\ll E_{\rm cr}$, where $\nu$ is the laser field frequency.  As criteria for the creation efficiency we use the mean density $\langle n\rangle$ per period and the residual density $n_r$, which is taken over an integer number of field periods \cite{Pop}.  Our main result is that optical lasers can generate a greater number of pairs than X-ray lasers in the spot volume $\lambda^3$.  This could be observable, e.g., through detection of coincident $\gamma$ pairs from electron-positron annihilation with mean total energy $\approx 1$ MeV.

\begin{figure}[t]
\centering
\resizebox{0.45\textwidth}{!}{%
\includegraphics{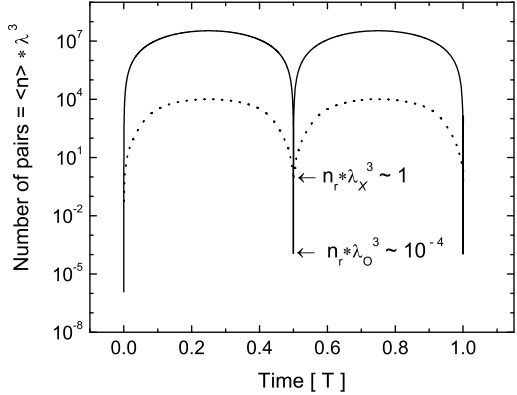}
}
\caption{\label{fig1:1} Time dependence of the number of produced e$^+$e$^-$ pairs in the volume $\lambda^3$ for a weak, periodic field, Eq.\,(\ref{harm}) $E_m = 3\cdot 10^{-5}E_{\rm cr} $, $\lambda = 795$ nm, corresponding to the Ti:sapphire laser \cite{Jena} (solid line) and for the near-critical field case of an X-ray laser \cite{Ring} with $E_m = 0.24 E_{\rm cr}$,  $\lambda =0.15$nm (dotted line).  The values of the residual pair density $n_r$ are marked in both cases.} 
\end{figure}

We consider here a simple model for the laser field that can be formed in the focus of two counter-propagating laser beams: an harmonic field that acts during $z$ periods {\bf $T=2\pi/\nu$}
\begin{equation}\label{harm}
E(t) = E_m \sin{\nu t},\qquad 0 \le t \le z T~,
\end{equation}
and vanishes for $t<0$ and $t>zT$.

Fig. \ref{fig1:1} depicts the time dependence of the quasiparticle pair density, which is generated by the field considered. The efficiency of plasma production is shown for the examples of a working optical Ti:sapphire laser \cite{Jena}, with $E_m \approx
3\cdot 10^{-5} E_{\rm cr} $,  $\lambda = 795$ nm, and for the planned X-FEL at DESY \cite{Ring}, with $E_m = 0.24 E_{\rm cr}$, $\lambda = 0.15$ nm. 
The density of e$^+$e$^-$ pairs oscillates with twice the field frequency $\nu$.  The residual density $n_r$, which corresponds to an integer number $z$ of field periods, $n_r=n(zT)$, is negligible in comparison with the mean density $\langle n\rangle$ for optical lasers.  The ratio of mean and residual densities is well approximated by the relation
\begin{equation}\label{nr}
\langle n\rangle  \sim \left(\frac{m}{\nu}\right)^2\, n_r \sim
\frac{(eE_m)^2}{m}  \, .
\end{equation}
For the case considered \cite{Jena} $\langle n\rangle \sim 10^7 \lambda^{-3}$ and the ratio $\langle n\rangle /n_r$ is approximately $ 10^{11}$ for the field in Eq.\,(\ref{harm}). As a consequence, despite the fact that the residual density for
the X-ray laser exceeds that of the optical laser by a large factor, the situation is different regarding the mean density: the optical laser produces more pairs in the spot volume than the X-ray laser.  According to Fig. \ref{fig1:1}, on average there are roughly $10^{7}$ pairs in a volume $\lambda^3 $, which corresponds to a pair density of $\sim 10^{20}$ cm$^{-3}$.  This estimate is in agreement with Refs.\,\cite{Sh,Av02}.  The dense plasma exists for the duration of a laser pulse but vanishes almost completely after switching off the field.

This result differs from that obtained with the imaginary time method \cite{Pop} where $n(z T)\sim z$.  That approach does not give information about $\langle n\rangle$.  We obtain that the mean density of electron-positron pairs is defined only by the field amplitude and does not depend on the frequency while $n_r \sim \nu^2$.  Both quantities are proportional to the intensity of the laser radiation.  This results in an accumulation effect for $n_r$ in the near-critical field of an X-FEL \cite{Rob}.  After an integer period number, most of the pairs vanish and the residual density becomes negligible in comparison with the mean one. The formula in Eq.\,(\ref{nr}) is inapplicable for a pulse shape field, which is a more realistic model of a laser beam.  In this case the behaviour of $n_r$ will depend strongly on the parameters describing the  pulse shape.   NB.\ This is not of concern when solving the equations numerically.

\begin{figure}[t]
\centering
\resizebox{0.4\textwidth}{!}{%
\includegraphics{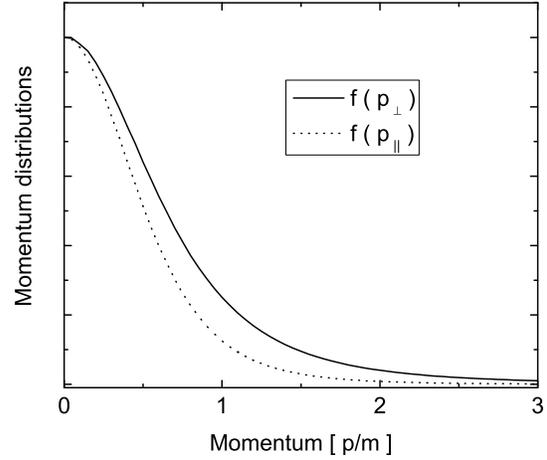}
}
\caption{Momentum distribution (in representative scale) of
created e$^+$e$^-$ pairs: p stands for transversal momentum (solid
line) and on longitudinal momentum (dotted line).} \label{fig2}
\end{figure}

\begin{figure}[t]
\centering
\resizebox{0.4\textwidth}{!}{%
\includegraphics{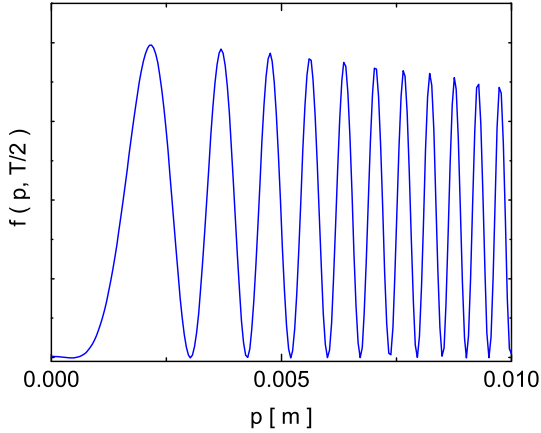}
}
\caption{Layered structure of momentum distribution ($p=|\mathbf{p}|$) at the time of zero field amplitude $t = zT/2$, associated with the residual density $n_r$ (in representative scale).} \label{lab}
\end{figure}

The momentum spectrum of created quasiparticle pairs is depicted in Figs.\,\ref{fig2}, \ref{lab}.  In contrast to the standard assumption of zero longitudinal momentum for e$^+$-e$^-$ pairs \cite{Casher}, the momentum distribution of the quasipartical pairs
has a width of the order of $m$ for both transverse and longitudinal momenta.  The shape of the momentum distribution is changed drastically when the electric field takes a zero value, Fig.\,\ref{lab}: a layered structure is formed with a characteristic time-scale corresponding to the field frequency.  This peculiar momentum distribution is associated with the residual density of e$^+$-e$^-$ pairs.

The quasiparticle plasma created in the weak field case can be manifested in various physical effects, such as nonlinear Thomson scattering \cite{Esarey93}, damping of electromagnetic waves \cite{BulanovSS}, one and two-photon annihilation \cite{Landau4}, the non-linear Breit-Wheeler process \cite{Ivanov}, etc.  
As an example, we estimate the intensity of two-photon annihilation in the plasma volume. The corresponding $\gamma$-quanta with mean total energy $\approx 1$ MeV can be
detected outside the focus of the counter-propagating laser beams. The production rate for this process is
\begin{multline}\label{num}
\frac{dN}{dV dt} = \int d \mathbf{p}_1 d \mathbf{p_2}\,
\sigma(\mathbf{p}_1,\mathbf{p}_2)f_1(\mathbf{p}_1,t)f_2(\mathbf{p}_2,t)\\
\times \sqrt{(\mathbf{v}_1 -\mathbf{v}_2)^2 - |\mathbf{v}_1 \times
\mathbf{v}_2 |^2} ,
\end{multline}
where $\mathbf{v}$ is the particle velocity and $\sigma$ is the
cross-section for two-photon annihilation \cite{Landau4}
\begin{multline}\label{sigma}
\sigma(\mathbf{p}_1,\mathbf{p}_2) = \frac{\pi e^4}{2m^2 \tau^2
(\tau-1)} \biggl[ \bigl(\tau^2 + \tau - 1/2 \bigr)\\
 \times \ln{\left\{ \frac{\sqrt{\tau} +
\sqrt{\tau-1}}{\sqrt{\tau} - \sqrt{\tau-1}}\right\} - (\tau+1)
\sqrt{\tau (\tau-1)} } \biggr].
\end{multline}
The t-channel kinematic invariant $\tau$ is given by
\begin{equation}\label{tau}
\tau = \frac{(p_1 + p_2)^2}{4m^2} = \frac{1}{4m^2}\bigl[
(\varepsilon_1 + \varepsilon_2)^2 - (\mathbf{p}_1 +
\mathbf{p}_2)^2 \bigr].
\end{equation}

\begin{figure}[t]
\centering
\resizebox{0.4\textwidth}{!}{%
\includegraphics{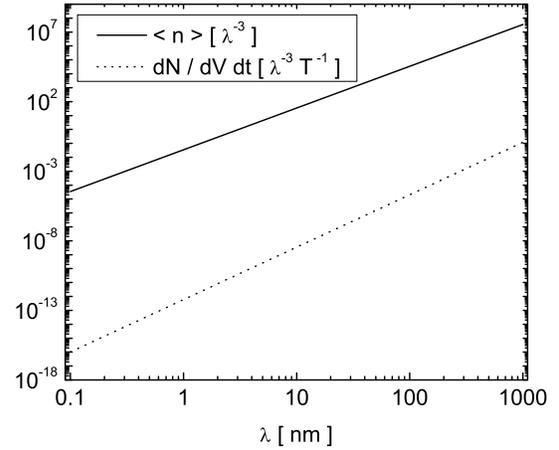}
}
\caption{ \label{fig4} Wavelength dependence of the mean density $\langle n \rangle$ (solid line) and the annihilation rate (dotted line) according to Eq.\,(\ref{num}) for a field strength $E= 3\cdot 10^{-5} E_{\rm cr}$.}
\end{figure}

We have estimated the number of annihilation events per laser pulse via Eq.\,\re{num} with the following parameters: pulse intensity $I=10^{20}$ W/cm$^2$, pulse duration $\tau_L \sim 85$ fs, wavelength $\lambda=795$ nm, crossing diameter of laser beams $\approx 2.5~\mu$m \cite{Jena}.  The result is approximately $5-10$ annihilation events per laser pulse. The wavelength dependence of the quantities discussed is shown in
Fig.\,\ref{fig4}.

We have argued that the simplest model of the laser field predicts the creation of a dense {quasiparticle} plasma in the foci of counter-propagating optical laser beams with
parameters corresponding operating lasers \cite{Jena,SLAC}.  The plasma lives during the laser pulse and vanishes almost completely after switching off the field.  The mean density is defined by the field strength and does not depend on frequency, reaching values $\sim ~10^{20}$ cm$^{-3}$ for a realistic field strength of $10^{11}$ V/cm.  A manifestation of the plasma may be the emission of pairs of $\gamma$-quanta with a spectrum peaked in the vicinity of a total energy of $1$ MeV, with an intensity of $5-10$ events per laser pulse.  This would be a non-linear transformation of the soft laser photons to $\gamma$-quanta with a frequency ratio of about $10^6$.

\subsection{Refraction experiment \label{sect:3:refr}}
We consider the possibility of an experimental verification of vacuum pair creation in the focus of two counter-propagating optical laser pulses (``photon collider'', \cite{PRL06,Heinzl_06}) by means of an interference experiment with an additional ultraviolet probe laser beam propagating across the plasma region.  We suppose that a short-lived electron-positron plasma can be created in the focus of the photon collider and that the plasma density is sufficiently large in order for its Langmuir frequency to become comparable with the optical one.  The probe laser beam must suffer refraction in this nonlinear optical medium, which could be detected by an interference refractometer.
We perform estimates for conditions that apply to the recently constructed petawatt laser Astra-Gemini, with an intensity $5\times 10^{21}$ W/cm$^2$ \cite{Astra}.  The effect turns out to be tiny and observation would be challenging.

Experiments using a short wavelength probe laser to diagnose the state created in the focus of colliding high-intensity laser pulses are discussed extensively in the literature  \cite{Heinzl_06,Heinzl_08,Piazza06}.  The polarized QED vacuum created in the ``photon collider'' acts like a birefringent medium, with two indices of refraction depending on the polarization of the incoming light.  This effect, discussed for the first time in Ref. \cite{KleinNigam}, was subsequently investigated in detail \cite{Baier67,Brezin71,Birula70}.  In Ref.\,\cite{Heinzl_06} the resulting ellipticity signal was calculated, which results from a phase retardation of one of the polarisation directions when a linearly polarized x-ray pulse passes through a hot spot generated by
high-intensity laser beams.  It was shown via the Heisenberg-Euler Lagrangian \cite{HE}
that the ellipticity signal is $\sim 10^{-13}$ for conditions of the Polaris laser \cite{Jena} with a probe frequency $~1$ keV, which corresponds to the limit of accuracy that can presently be obtained with high-contrast X-ray polarimeters.  This scheme is applied to the case of the polarized vacuum generated by two counter-propagating beams in Ref.\,\cite{Piazza06}.  In that case the probe signal is directed across the high-power beam so that, in principle, a refraction of the probe beam can be expected.  The rotation of polarization is estimated to be on the order of $10^{-8}$ rad.

\begin{figure}[t]
\centering
\resizebox{0.45\textwidth}{!}{%
\includegraphics{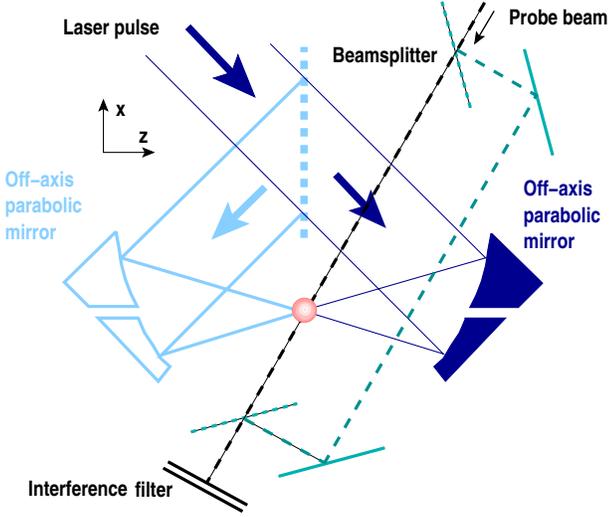}
}
\caption{Setup for a measurement of the refraction index by the probe laser
interference method.  The short-wavelength probe laser beam $\lambda_p \ll \lambda_0$ propagates across the direction of the high-power beams. \label{fig:interf}}
\end{figure}

Unlike schemes designed to measure the polarization of the vacuum, we suggest to consider the interaction of a probe laser pulse directly with the pulsating $e^+e^-$-plasma forming a peculiar spatial ``diffraction grating'' pattern or a one-dimensional photon crystal.  In the focus of two counter-propagating high-power laser beams, the standing
wave appears for a short time, so that the Schwinger mechanism starts to work close to the wave antinodes whereat the plasma ``grating'' is formed.  The refraction of one of two split probe beams can be measured by the usual interference method, see Fig. \ref{fig:interf}.  The size of the plasma region is of the order of the optical wave length and its density pulses with twice the optical frequency, so that a stable
interference pattern can be obtained with a sufficiently short probe signal; i.e, $\tau_p \sim \lambda_0/2$.  We use for numerical estimates the parameters of the petawatt Astra-Gemini laser \cite{Astra}: $\lambda_0 = 800$ nm, 
$I = 5\times 10^{21}$ W/cm$^2$.

Detection of the reflected beam would be technically easier but the reflection is strongly suppressed in this case owing to the absence of a sharp plasma boundary.  The characteristic scale of inhomogeneity, $\Delta L$, in the probe beam direction is defined by the cross section of the laser focus; viz., $\lambda_p \ll \Delta L$.  This justifies the application of geometrical optics when reflection can be neglected \cite{Landau8}.

Owing to the dependence of the medium's properties ($e^+e^-$-plasma) on the electric field of the basic laser, the calculation of the refraction index of the probe signal requires nonlinear optics.  However, the mode decomposition of the field which is traditional for such problems cannot be used in our case because the effect of pair creation is a truly nonperturbative one.  We therefore consider the weak probe field in linear approximation as the basis for an exact solution for the high-power laser field.

To account for the influence of the basic laser's strong field on the dielectric properties of the plasma generated by it, we use a kinetic equation in the form of Eqs.\,(\ref{ode}).  We choose the geometry of the probe field as $\mathbf{k} = (k,0,0)$, which is convenient since it is then possible to consider the fields of the basic and the probe laser as collinear and directed along the axis OZ.
%so that
%\begin{equation}
%\label{refr:8}
%\begin{split}
%\varepsilon_\perp & = \sqrt{m^2 + q^2_{\trans}},\\[3pt]
%\varepsilon &= \sqrt{\varepsilon^2_{\trans}+(q_{\spar}-eA)^2}, \\[3pt]
%\Delta & = eE \ \varepsilon_\perp / \varepsilon^2, \\[3pt]
%E(t) &= - \dot{A}(t). \end{split}
%\end{equation}
All quantities are decomposed in a standard way \cite{Landau10},
\begin{eqnarray}\label{refr:9}
f = f_0 + \delta f,\quad u &= &u_0 + \delta u,\quad v = w_0 + \delta w,  
\nonumber \\[4pt]
\mathbf{E} &=& \mathbf{E}_0 + \delta \mathbf{E}, \nonumber \\[4pt]
\delta f, \delta u, \delta w, \delta \mathbf{E} &\sim & \exp{(-i\var t + i
\mathbf{k} \mathbf{r})}
\end{eqnarray}
but with an essential difference that the functions $f_0, u_0, w_0$ are the
exact solutions of the kinetic equation in the strong field $\mathbf{E}_0$,
instead of equilibrium ones as in plasma theory.

To describe spatial dispersion we introduce an essential assumption that the kinetic equation, Eq.\,\re{ke}, is applicable for the description of small inhomogeneous perturbations; namely, we add the standard operators $\mathbf{v}(\partial /\partial \mathbf{r})$ to the left hand side of the equations.  We now perform the substitution
\begin{equation}
\label{refr:10}
\mathbf{p} = \mathbf{q} - e\, \delta\mathbf{A},
\end{equation}
i.e., effect a transition to the kinematic momentum related to the perturbation
field $\delta \mathbf{E}$, then
\begin{eqnarray}
\label{refr:11}
    \frac{\partial f}{\partial t} 
+ \mathbf{v}\frac{\partial f}{\partial \mathbf{r}} 
+ e\,\delta \mathbf{E} \frac{\partial f}{\partial \mathbf{p}} 
&=& \frac12\, \Delta u, \nonumber\\
\frac{\partial u}{\partial t} 
+ \mathbf{v}\frac{\partial u}{\partial \mathbf{r}} 
+ e\,\delta \mathbf{E} \frac{\partial u}{\partial \mathbf{p}} 
&=& \Delta\, (1- 2 f) - 2 \varepsilon_0\, w, \\
\frac{\partial w}{\partial t} 
+ \mathbf{v}\frac{\partial w}{\partial \mathbf{r}} 
+ e\,\delta \mathbf{E} \frac{\partial w}{\partial \mathbf{p}}
&=& 2 \varepsilon_0\, u , \nonumber
\end{eqnarray}
where $\mathbf{v} = \mathbf{p}/\var_0$.

We substitute here the decomposition in Eq.\,\re{refr:9} and compute the first order corrections
necessary for a calculation of the current.
We can then determine the current density, Eq.\,\re{current},
\begin{equation}\label{refr:19}
  \delta j = \frac{e}{2\pi^3} \int \frac{d \ve{p}}{\varepsilon_0} \left[ p_{\along}
    \delta f + \frac12 \varepsilon_\perp \delta u \right],
\end{equation}
and define a permittivity with the help of the relations
\begin{equation}\label{refr:20}
    \begin{split} j_{\alpha} &= \sigma_{\alpha\beta}E_{\beta}, \\[5pt]
    \epsilon_{\alpha\beta} &= \delta_{\alpha\beta} + \frac{4\pi i}{\omega}
     \sigma_{\alpha\beta}. \end{split}
\end{equation}

Substituting the regularized $\delta f, \delta u$, we obtain for the
transverse part of the permittivity
\begin{multline}
\label{refr:21}
\epsilon_{tr} = 
1 - \frac{2e^2}{\pi^2 \omega  } \int \frac{d^3 p}
{\var_0 (\Omega^2 - \Delta_0^2)} 
\left\{ \frac{P_{\along} \Omega^2}{(\omega - \ve{k}\ve{v})} 
\left[ \frac{\varp}{2 \var_0^2} u_r - \frac{\partial f_r}{\partial p_{\along}} 
\right. \right. \\
- \left. \left. \frac{\varepsilon_0 \Delta_0}{\Omega^2} \frac{\partial w_r}{\partial
p_{\along}}\right] \right.\\[5pt]
\left. + \varp (\omega - \ve{k}\ve{v}) \left( \frac{\varp}{8\var^4_0} \biggl[(\omega - \ve{k}\ve{v})^2 - \Delta_0^2 \biggr] - \frac{\varp}{\var_0^2}\, f_r - \frac12 \frac{\partial u_r}{\partial p_{\along}} \right)
\right\}  \\[5pt]
+ i \frac{e^2}{\pi^2 \omega } \int \frac{d^3 p}{ \var_0 (\Omega^2 -
\Delta_0^2)} \left\{  P_{\along} \Delta_0 \left[ \frac{\partial u_r}{\partial p_{\along}}
- \frac{\varp}{\var_0^2} (1 -2 f_r) \right] \right. \\ \left. + 2 \varp \left[ \Delta_0 \left(
\frac{\varp}{2 \var_0^2} u_r - \frac{\partial f_r}{\partial p_{\along}} \right) - \var_0
\frac{\partial w_0}{\partial p_{\along}} \right] \right\},
\end{multline}
where
\begin{equation}\label{refr:15}\begin{split}
\Omega^2 = (\omega -   \mathbf{k}\mathbf{v})^2 - 4 \varepsilon_0^2, \\[3pt]
\Delta_0 = e E_{ex}(t) \varepsilon_\perp / \varepsilon_0^2.
 \end{split}
\end{equation}

Unlike the non-relativistic case \cite{Landau10}, the denominator $(\omega -
\ve{k}\ve{v})$ does not generate a pole for all $\omega, \ve{k}$, but only for
${k/\omega} > 1$.  Since our model does not allow an analytic calculation of $\epsilon_{tr}$ for the whole range of variables, we take early advantage of the dispersion relation for transverse fluctuations in the oscillations at the stage
of calculating the momentum integral in Eq.,\,\re{refr:21}; i.e., we exploit the condition
\begin{equation}\label{refr:22}
\omega^2 \epsilon_{tr} - k^2 = 0~.
\end{equation}
We therefore write $\mathbf{k} = (\omega \sqrt{\epsilon_{tr}},0,0)$ and solve
the equation by iteration with respect to $\epsilon_{tr}$, choosing $\epsilon^{(0)}_{tr}=1$ as the initial point.  Since the pole is absent in this approximation, we can integrate over the polar angle, taking into account that the optical frequency satisfies $\omega\ll m $ and hence $(\Omega^2 - \Delta_0^2) \approx -4 \var^2$ for $E\ll E_{\rm cr}$.  On the same basis it is possible to neglect the second string in the real part of Eq.\,\eqref{refr:21}.  Limiting ourselves to the case of a non-degenerate plasma: $f_0 \ll 1$, we have
\begin{multline}
\label{refr:23}
\epsilon^{(1)}_{tr} = 
1 - \frac{2 e^2}{\pi^2 \omega^2} \int \  d^3 p\ \frac{P_{\along}}{\vara}\ 
\left[ \frac{\varp}{2 \var_0^2} u_r - \frac{\partial f_r}{\partial p_{\along}} 
+ \frac{\Delta_0}{4 \var_0} \frac{\partial w_r}{\partial p_{\along}}\right] 
\\[5pt]
- i \frac{e^2}{4 \pi^2 \omega } \int \frac{d^3 p}{ \var_0^3} \left\{
P_{\along} \Delta_0 \left[ \frac{\partial u_r}{\partial p_{\along}} -
\frac{\varp}{\var_0^2} \right] \right. \\ 
\left. + 2 \varp \left[\Delta_0  \frac{\varp}{2 \var_0^2} u_r 
- \var_0 \frac{\partial w_r}{\partial p_{\along}} \right] \right\} .
\end{multline}

\begin{figure}[t]
\centering
\resizebox{0.45\textwidth}{!}{%
\includegraphics{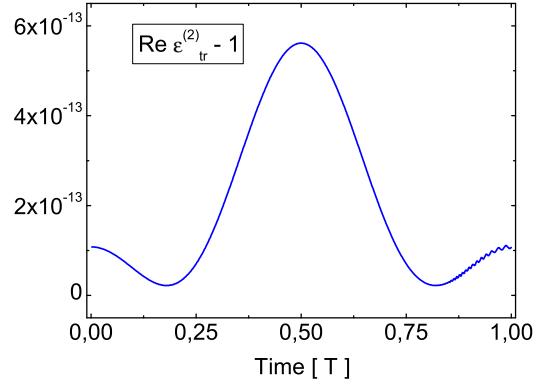}
}
\caption{The real part of the transverse permittivity calculated in second 
order of iterations \re{refr:21}.} \label{fig:6}
\end{figure}

\begin{figure}[t]
\centering
\resizebox{0.45\textwidth}{!}{%
\includegraphics{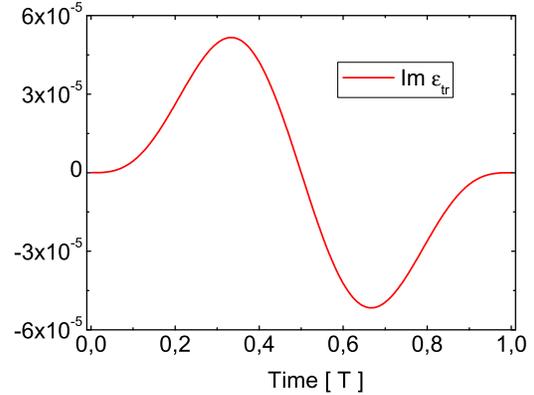}
}
\caption{The imaginary  part of the transverse permittivity} \label{fig:7}
\end{figure}

The numerical evaluation of Eq. \eqref{refr:23} shows that $\epsilon^{(1)}_{tr} > 1 $.
Therefore the calculation of the next iteration in Eq.\,\re{refr:21} must be done by respecting a pole at $ v^2_x =1/\epsilon^{(1)}_{tr}$ using the known Landau recipe (Dirac identity) \cite{Landau10}.
The contribution of the $\delta$-function to the integral is localized in the
region of very large momentum $p \gg m $ and can therefore be neglected.  The principal value of the integral is calculated at leading order of the expansion in the small parameters $\omega/m$ and $\Delta_0/\var_0$:
\begin{multline}
\label{refr:25}
\mathrm{Re}\,\epsilon^{(2)}_{tr} = 1 - \frac{e^2}{(2\pi)^2} \int \ \frac{d^3
p}{\var_0^3} \left\{ \frac{2 \varp^2}{\var_0^2} f_r + \varp \frac{\partial u_r}{\partial
p_{\along}}\ \right. \\ \left. + \frac{\Delta_0 P_{\along}}{ 2\var_0}\frac{\partial w_r}{\partial
p_{\along}} \right. \\ \left. + \left( \frac{\varp }{2 \var_0^2} \right)^2
\left[\Delta_0^2 - \omega^2 \left(1+ \frac{3\, p_\perp}{2\, \var_0} \right) \right] \right\}~.
\end{multline}

Figs.\,\ref{fig:6}, \ref{fig:7} show that the variation of the refraction index of a probe beam in comparison with the vacuum case is tiny.  It poses a challenge for experimental observation, at least for the geometry considered.  Other field geometries need separate consideration owing to the need for use of more complex kinetic equations \cite{Filatov,Skokov}, supposing an electric field of variable direction.  The conclusion obtained is in accord with previous estimates \cite{Heinzl_06,Piazza06}.

\section{Extensions towards the Schwinger limit \label{sec:4}}
\subsection{Application in the near-critical case (XFEL) %\cite{Rob}
\label{sec:4:1}}
A kinetic equation coupled with Maxwell's equation was used to estimate the laser power required at an XFEL facility to expose the process of QED vacuum decay via spontaneous pair production \cite{Rob}.  A $9\,$TW-peak XFEL laser with photon energy $8.3\,$keV could be sufficient to initiate particle accumulation and the consequent formation of a plasma of spontaneously produced pairs.  The evolution of the particle number in the plasma will exhibit non-Markovian aspects of the strong-field pair production process and the plasma's internal currents will generate an electric field whose interference with that of the laser leads to plasma oscillations.

X-ray free electron laser (XFEL) facilities are planned at SLAC~\cite{SLAC}, namely the Linac Coherent Light Source (LCLS) \cite{LCLS}, and the XFEL at DESY which emerged from the design studies for the $e^-e^+$ linear collider project (formerly TESLA now ILC: International Linear Collider) \cite{desy}.  They propose to provide narrow bandwidth, high power, short-length laser X-ray pulses, with good spatial coherence and tunable energy.  It is anticipated that the realizable values of these parameters will enable studies of completely new fields in X-ray science, with applications in atomic and molecular physics, plasma physics and many other fields \cite{desy}.

A unique ability of these facilities is to provide very high peak power densities.  For example, a $P=0.2\,$TW-peak laser at a wavelength of $\lambda=0.4\,$nm, values which are reckoned achievable with current technology \cite{desy}, can conceivably produce a peak electric field strength
\begin{equation}
E_a = \sqrt{ \mu_0 c\, P/(\pi \lambda^2)} \approx 1.2\times 10^{16} \, {\rm
V/m}\,.
\end{equation}
Boosting $P$ to $1$~TW and reducing $\lambda$ to $0.1\,$nm, which is theoretically possible \cite{Chen}, would yield an order-of-magnitude increase: $E_g = 1.1 \times 10^{17}$~V/m.  Electric fields of this strength are sufficient for an experimental verification of the spontaneous decay of the QED vacuum \cite{Ring,Fried,Popov01,bastiprl}.

A single laser beam cannot produce pairs \cite{Troup}.  (For a light-like field $F_{\mu\nu} F^{\mu\nu} = 0$ and hence the vacuum survival probability is equal to one.)
Nevertheless, if two or more coherent beams are crossed and form a standing wave at their intersection, one can hypothetically produce a region in which there is a strong electric field but no magnetic field.  The radius of this spot volume is diffraction limited to be larger than the laser beams' wavelength: $r_\sigma \gtrsim \lambda$, and the interior electric field could be approximately constant on length-scales approaching this magnitude.  The period of the electric field is also determined by $\lambda$.  Hence at an XFEL facility one might satisfy the length-scale uniformity conditions noted above.

However, the planned laser pulse duration: $\tau_{\rm coh} \sim 100\,$fs, is large compared to the laser period: $\tau_\gamma \sim 1\,$as, and thus the time-dependence of the electric field in the standing wave may materially affect the pattern of observed pair production; i.e., vacuum decay might be expressed via time-dependent pair production.  This possibility can only be explored using methods of non-equilibrium quantum field theory \cite{Brezin70,Popov72,NN73} or an equivalent quantum kinetic
theory \cite{Smol,our,kme}.  The latter procedure was employed in Ref.\,\cite{bastiprl}, wherein it was shown that pair production occurs in cycles that proceed in tune with the laser frequency.  While that does not lead to significant particle accumulation, the peak density of produced pairs is frequency independent, with the consequence that several hundred pairs could be produced per laser period.

\begin{table}[b]
\caption{\label{table} Laser field parameters are specified in columns one and two: Set~I is XFEL-like; Set~II is strong.  Columns three and four describe the density, $n_{max}(t_>)$, and total number of produced particles, $N(t_>) = \lambda^3 n_{max}(t_>)$, where $t_>$ is the time at which the number density reaches its (for weak fields, local) maximum.  A typical laser pulse length is $\sim 80\,$fs \protect\cite{desy}.}
\vspace*{1ex}
\begin{tabular*}
{\hsize} {l@{\extracolsep{0ptplus1fil}}
|l@{\extracolsep{0ptplus1fil}}l@{\extracolsep{0ptplus1fil}}
|l@{\extracolsep{0ptplus1fil}}r}
%{l|c|c|c}
          & $\lambda\,$(nm) & $E_0\,$(V/m) & $n_{max}(t_>)\,$(fm$^{-3})$
          &$N(t_>)$ \\\hline
%         & & & & \\
Set ~Ia    & 0.15     & $1.3\times 10^{17}$
&$4.6\times 10^{-13}$ & $\sim 10^3$ \\
Set ~Ib    & 0.075     & $1.3\times 10^{17}$
&$4.6\times 10^{-13}$ &$\sim 10^2$\\
Set IIa   &  0.15    & $1.3\times 10^{18}$
&$7.2\times 10^{-10}$ & $\sim 10^{6}$  \\
Set IIb   &  0.075    & $1.3\times 10^{18}$
&$6.4\times 10^{-10}$ &$\sim 10^{5}$\\
\end{tabular*}
\end{table}

\begin{figure}[t]
\centering
\resizebox{0.45\textwidth}{!}{%
\includegraphics{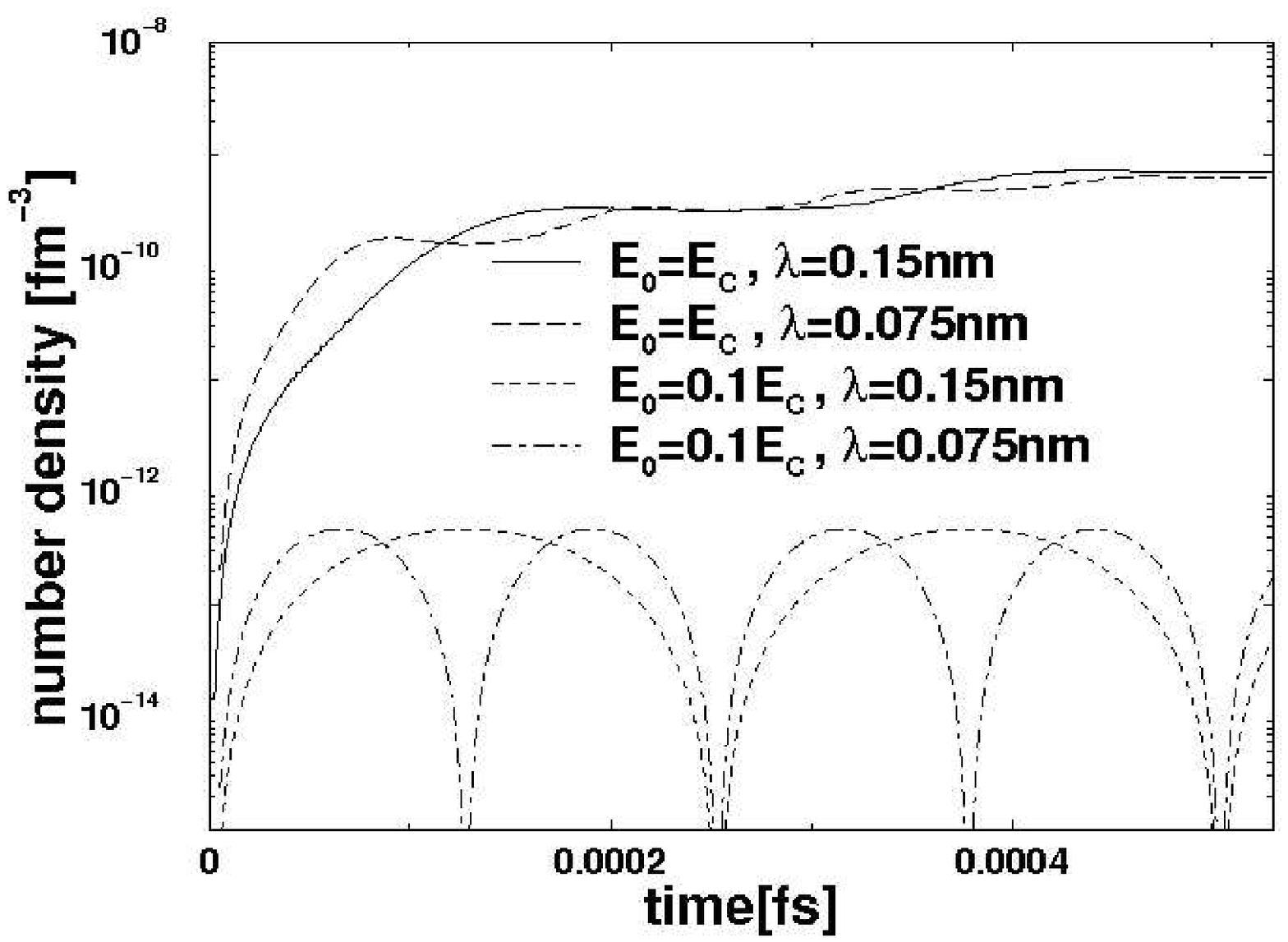}}
\caption{\label{fig:x1} Time evolution of the number density for Sets I and II, Table~\protect\ref{table}.  In strong fields, particles accumulate, leading to the almost complete occupation of available momentum states.  In weak fields, repeated cycles of particle creation and annihilation occur in tune with the laser frequency: $\Omega_{\rm Ia}=2.0 \times 10^{18}\,2\pi s^{-1}$; $\Omega_{\rm Ib}=4.0 \times 10^{18}\,2\pi s^{-1}$.}
\end{figure}

\begin{figure}[t]
\centering
\resizebox{0.45\textwidth}{!}{%
\includegraphics{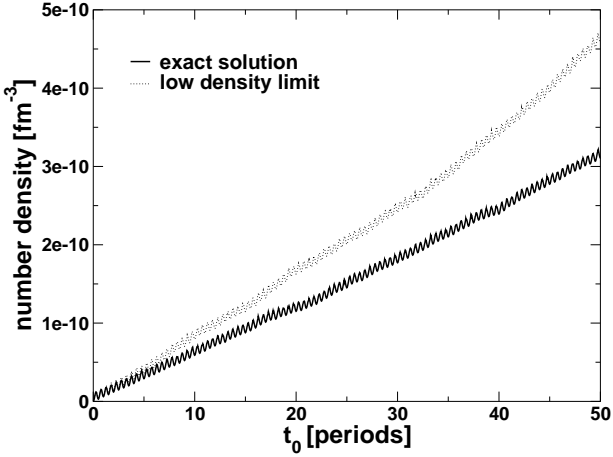}} \caption{\label{fig:x2}
Number density calculated with $E_0=0.5\,E_{\rm cr}$.  Solid line: solution of
Eqs.~(\ref{ke}), (\ref{max}); dotted line: solution obtained
using a low density approximation Eq.~(\ref{fact}).  The
oscillations are tied to the laser frequency, see Eq.~(\ref{ntfit}).}
\end{figure}

The proposed XFEL facilities offer a first real chance of observing the decay of QED's vacuum, a profound and nonperturbative quantum field theoretical effect.  Nevertheless, with a quantum Vlasov equation, one can ask for more.  This equation yields the time-dependence of the single particle distribution function: $f(\vec{p},t):= \langle
a^\dagger_{\vec{p}}(t)\,a_{\vec{p}}(t) \rangle$, and hence can be used to estimate the laser power required to achieve an accumulation of $e^-\,e^+$ pairs via vacuum decay.  Furthermore, in quantum field theory the particle production process is necessarily non-local in time; i.e., \textit{non-Markovian}, and dependent on the particles' statistics.  These features are preserved by the source term in the Vlasov equation. (The Schwinger source term is recovered in a carefully controlled weak-field limit~\cite{kme}.)  Consequently, one can identify the laser parameters necessary to expose negative energy elements in the particle wave packets.  The system exhibits this core quantum field theoretical feature when the time between production events is commensurate with the electron's Compton wavelength.  Along with accumulation comes the possibility of plasma oscillations, generated by feedback between the laser-produced electric field and the field associated with the production and motion of the $e^-\,e^+$ pairs; and also collisions.  (These features are reviewed in Ref.~\cite{bastirev}.)

We model an ``ideal experiment,'' and assume $r_\sigma = \lambda$ and a nonzero
electric field constant throughout this volume while the magnetic field vanishes identically.  This is impossible to achieve in practice and therefore the field strengths actually achievable will be weaker than we suppose.  Hence our estimates of the laser parameters will be lower bounds. Our model is represented by the set of equations (\ref{ke}),(\ref{max}) and (\ref{harm}).  (Table~\ref{table} provides our laser field parameters.)  There are two control parameters in Eq.~(\ref{ke}): the laser field strength, $E_0$, and the wavelength, $\lambda$.  We fix $\lambda = 0.15\,$nm, which is
achievable at the proposed XFELs.  (NB.\ By assumption, the volume in which particles are produced increases with $\lambda^3$ whereas the field strength decreases with $1/\lambda$: there is merit in optimising $\lambda$.) Our study will expose additional phenomena that become observable with increasing $E_0$.

\begin{figure}[t]
\centering
\resizebox{0.45\textwidth}{!}{%
\includegraphics{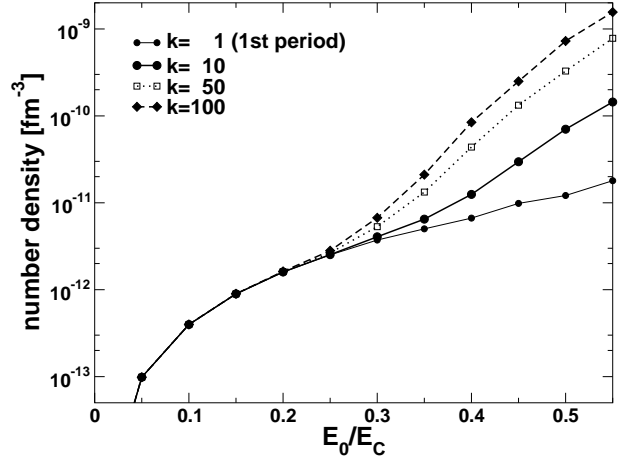}} \caption{\label{fig:x3}  Peak particle number density versus laser field strength. There is a striking  qualitative change at $E_0 \approx 0.25\,E_{\rm cr}$, which marks the onset of particle accumulation.}
\end{figure}

The quantitative analysis was performed for two exemplary electric field strengths, the weaker field $E=0.1\,E_{\rm cr}$, which should be obtainable at the proposed XFEL facilities \cite{SLAC,desy} and $E=E_{\rm cr}$, to provide a strong field comparison.  The results are presented in Table~\ref{table} and Fig.~\ref{fig:x1}.  The dynamics of pair creation is qualitatively changed at field strengths $\sim 0.5 E_{\rm cr}$.  The detailed investigation of this feature is presented in Fig,~\ref{fig:x2}.  The results for $T=t/\lambda \lesssim 100$ are accurately fitted by
\begin{eqnarray}
\label{ntfit}
n(T;E_0) & = & a_0(E_0) \, \sin^2 2\pi T\, + \rho(T,E_0)\,T \,, \\
\label{kappadef}
\rho(T,E_0) & = & \rho(E_0) + \rho^\prime(E_0)\, T \,.
\end{eqnarray}
One can therefore use $\rho(T,E_0)$ to quantify the rate of particle accumulation.  This rate is very small for $E_0 \lesssim 0.2\,E_{\rm cr}$ \cite{bastiprl} and, while noise in $n(t)$ prevents a reliable numerical determination of $\rho$, $\rho^\prime$, it is nevertheless clear that in this case lengthening $\tau_{\rm coh}$ will not materially increase the number of particles produced.  It is apparent from Fig.\,\ref{fig:x2}, however, that the situation is very different for $E_0= 0.5\,E_{\rm cr}$.  The solid curve is described by Eq.~(\ref{ntfit}), with $a_0= 1.2 \times 10^{-11}\,$fm$^{-3}$, $\rho= 5.4 \times 10^{-12}\,$fm$^{-3}\,$period$^{-1}\!$, $\rho^\prime/\rho= 0.0033/$ period.  Clearly, the accumulation rate is approximately constant.

Figure~\ref{fig:x3} shows the \textit{peak} particle number density, $n(t_k^>)$, where $t_k^>= (4 k -3) \lambda/4$ is the time at which $E(t)$ is maximal during laser period no.\ $k$.  The figure displays a qualitative change in the rate of particle production with increasing laser field strength.  ($n(k \lambda) \approx n(t_k^>)$ if and only if there is significant particle accumulation, otherwise $n(k \lambda)\ll n(t_k^>)$.)

For $E_0 \lesssim 0.25 \,E_{\rm cr}$ there is no significant accumulation of
particles: $N(t_0^>,E_0) \approx N(t_{10}^>,E_0)$, etc., just as observed in
Ref.\,\cite{bastiprl}.  However, for $E_0 > 0.25 \,E_{\rm cr}$ there are more
particle pairs after each successive laser period; e.g., $E_0 = 0.35\,E_{\rm
cr}$ brings an order of magnitude increase in $N$ over the first $100$ laser
periods.  At such values of $E_0$ one is in a domain where numerous electrons
and positrons produced in a single period are accelerated to relative
longitudinal momenta that are sufficient to materially inhibit annihilation.
This ensures that many pairs remain when the next burst of production occurs.
Consequently $N(t_k^>,E_0)$ grows considerably with increasing $k$.
Accumulation means collisions can become important and should be included in
the kinetic equation if quantitatively reliable results are required.

\subsection{Muon and pion pair creation \label{sec:4:2}}
The rapid advancement expected in laser technology, progressing toward the Schwinger limit case for electron-po\-si\-tron pair creation within the XFEL as well as the ELI project \cite{Paris}, stimulates the question about vacuum production of charged particles heavier than electrons; e.g., muons and pions.

The mean pair density per field period is inversely proportional to the particle mass, Eq.\,(\ref{nr}), so that the mean density of muon and pion pairs is estimated as $n_\mu = n_e/207$ and $n_\pi = n_e/273$.  In order to achieve for $\pi^\pm$ and $\mu^\pm$ pairs the same densities as for $e^\pm$ pairs, one has to increase the laser intensity by 2-3 orders of magnitude.  According to estimates in Sect. \ref{sec:2:1} for the annihilation
process of $e^\pm$, a rescaling for $\mu^\pm$ pairs will result in a few hard ($\sim 100$ MeV) $\gamma$-quanta per laser shot.  For the determination of the $2 \gamma$ rate from $\pi^\pm$ annihilation we use here Eq.~(\ref{sigma}) together with the vacuum cross-section for two-photon annihilation of pointlike scalar mesons \cite{AB}
\begin{multline}
\sigma(\mathbf{p}_1,\mathbf{p}_2) = \frac{e^4}{4\pi m^2 \tau ^2}
\left\{ \frac{(\tau +1)\sqrt{\tau}}{\sqrt{\tau -1}} \right. \\ \left. -\frac{(2\tau
-1)}{(\tau-1)} \ln{(\sqrt{\tau} + \sqrt{\tau -1})}\right\}, 
\end{multline}
where $\tau$ is given by Eq.~(\ref{tau}).

\begin{figure}[t]
\centering
\resizebox{0.45\textwidth}{!}{%
\includegraphics{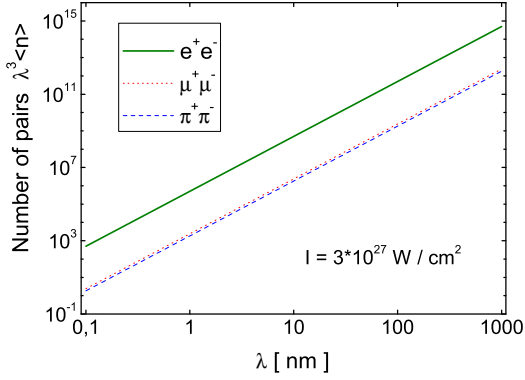}
}
\caption{Dependence of the mean number of particle-antiparticle pairs in the
volume $\lambda^3$ as a function of the wavelength for a periodic field with
intensity $I= 3\cdot 10^{27}$ W/cm$^2$ ($E_m = 10^{15}$ V/cm) for different
particle species: electrons (solid), muons (dotted) and pions (dashed).
\label{fig:mu1}}
\end{figure}

\begin{figure}[t]
\centering
\resizebox{0.45\textwidth}{!}{%
\includegraphics{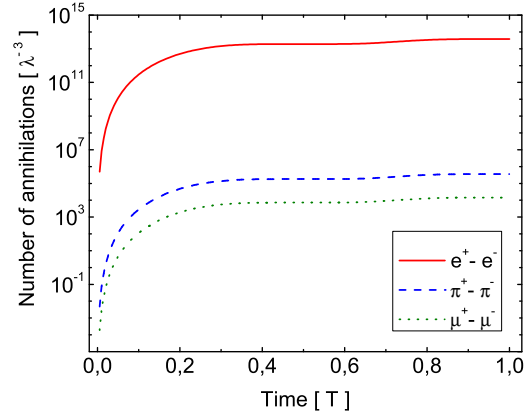}
}
\caption{Time dependence of the number of annihilations in the volume $\lambda^3$ for the same conditions as in Fig.~\ref{fig:mu1}.
\label{fig:mu2}}
\end{figure}

In Fig. \ref{fig:mu1} we show the dependence on wavelength of the mean pair densities and in Fig. \ref{fig:mu2} the number of two-photon annihilations as a function of time for $e^\pm$, $\pi^\pm$ and $\mu^\pm$ pairs at a field intensity of $I=3\times 10^{27}$ W/cm$^2$.

Here we used (for simplicity of numerical calculation) an illustrative model: a
harmonic laser field persisting for an integer number of periods.  As a characteristic quantity we show the density of created pairs in the node of the laser field, which is small but nonvanishing even for subcritical field strengths, see Fig.~\ref{fig:mu1}.
For a more realistic model, such as a laser pulse with Gaussian envelope, it can be shown that the number of residual pairs for subcritical field strengths is exponentially suppressed by the large factor $m \tau$, where $\tau$ is the pulse duration.  Therefore the only observable traces of the dynamical Schwinger effect in the focus of
optical lasers are the products of the secondary reactions of short-living quasiparticles as, e.g., the two-photon annihilation processes \cite{Piazza07,Nik64} or quasiparticle decays.

\begin{figure}[t]
\centering
\resizebox{0.45\textwidth}{!}{%
\includegraphics{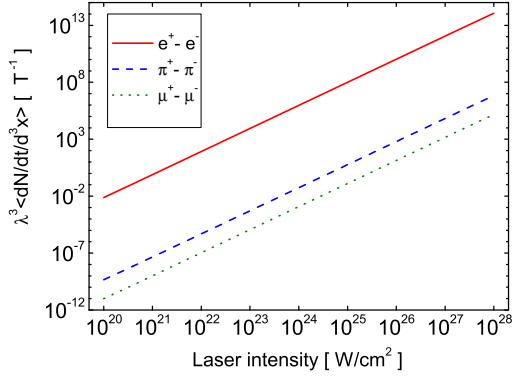}
}
\caption{Number of annihilations per unit volume and time as a function of the
laser intensity for $e^+e^-$ (solid), $\pi^+\pi^-$ (dashed) and $\mu^+\mu^-$
(dotted) pairs. \label{fig:13}}
\end{figure}

In Fig.~\ref{fig:13} we show the rate of 2$\gamma$ annihilation processes per
volume for  $e^+e^-$, $\pi^+\pi^-$ and $\mu^+\mu^-$ pairs as a function of the
laser intensity.  The  $e^+e^-$ annihilation into pairs of $511\,$keV photons
is clearly dominating, but may be distinguishable from the other processes because they generate hard $\gamma$-ray photons of about 100 MeV energy.  According to this result,
the rates from $e^+e^-$ annihilation at presently available laser intensities
of $10^{22}$ W/cm$^2$ \cite{Astra} would correpond to those for annihilation of $\pi^+\pi^-$ and $\mu^+\mu^-$ pairs at intensities of $10^{26}-10^{27}$ W/cm$^2$, planned within the ELI project \cite{ELI}.

Another method of muon pair creation is considered in Refs.\,\cite{Keitel_mu}.  It consists in the coherent collisions of $e^-$ and $e^+$ produced from a positronium atom by a strong laser pulse.  This mechanism is an alternative to the conventional one taking place in non-coherent $e^-e^+$  colliders.  In our case we have partly similar conditions when the created matter represents a ``condensate'' of pairs with zero momentum \cite{Grib}.  Mathematically this is expressed by the condition $f(\mathbf{p},t) =
\tilde{f}(-\mathbf{p},t)$, where $f$ and $\tilde{f}$ are the momentum distributions of particles and antiparticles.  In the approximation used this correlated state is conserved during the time evolution, which assists the reabsorption of all pairs when the electric field disappears.

\section{Heavy ion acceleration in vacuum with high-intensity laser beams 
\label{sec:5}}
We review here some of our investigations on the possibility of additional vacuum acceleration of heavy ions in laser fields of different configurations \cite{ILIAS07,ILIAS08}.  The main observation is the existence of a big variety of acceleration modes owing to many fitting parameters, even for only one beam of Gaussian shape.  For crossed beams there are even more options.  The relevant variables, such as initial velocity or pulse duration, produce an essentially non-monotonic dependence of the energy gain that makes the search for the most effective mode acceleration very complex.  A threshold level of intensity exists ($\sim 10^{24}$ W/cm$^2$) when the ion moves in the capture mode in one direction.  The crossed-beams scheme is not less than three times more effective than the one-beam scheme in the considered range of parameters. However, such a scheme works for a definite phasing of beams only, which is difficult to obtain at high field intensity.  In such a regime other non-linear effects (pair creation and vacuum polarization) can also be active.

The idea of particle acceleration by means of laser fields was first proposed in
Refs.\,\cite{Shimoda,Ashkin} and has been investigated carefully for a long time, with attention mainly to methods using plasma-beam interaction.  The laser wakefield
accelerator concept was introduced in Ref.,\cite{Tajima} and has become widespread in recent years.  It is now close to practical realization \cite{Malka,Patel}.  A variation of this method is the interaction of high-intensity laser pulses with solid targets, which has proven to be effective for applications such as proton radiography or isotope production and nuclear activation \cite{target}.

The rapid development expected for ``table-top'' high-intensity lasers \cite{Paris,Ren} opens the possibility for additional vacuum acceleration of heavy ion beams and has been discussed, e.g., in the context of the upgrade of existing accelerators, such as the Dubna nuclotron \cite{Dubna}.  In previous work \cite{ILIAS07} we proved that such a scheme should be realizable at a laser intensity of the order of $10^{25}-10^{26}\,$W/cm$^2$.  The presence of plenty of fitting parameters, upon which the result depends in an essentially non-monotonic fashion, makes it complicated to
find the most effective acceleration mode.  The authors of Ref.\,\cite{BT} have concluded that circular polarization is most effective for electron acceleration.  In contrast we have found that linear polarization is more preferable for heavy ion acceleration. Furthermore, vacuum pair creation, which is initiated at such high field intensity
\cite{Rob,Ring,PRL06},  can change considerably the conditions for ion beam motion.

Since a direct calculation of the creation rate for a Gaussian shaped field is too complex for the present status of the approach, we perform an estimate with the help of the field invariants.  The presence of short lived plasma matter can cause acceleration of particles in itself even without participation of the laser field, as was recently
demonstrated in the SLAC experiment \cite{Blumen}.  The correct description
of ion-laser beam interaction at intensities above $10^{22}$ W/cm$^2$ must account for non-linear vacuum effects as well as the plasma-beam interaction.

\subsection{One beam scheme}
Following Refs.\,\cite{BT,Salam2006} we consider heavy ion acceleration as the motion of classical point particles in the prescribed electromagnetic field
\begin{align}
\dot{\mathbf{p}}&= e \bigl( \mathbf{E} +  \mathbf{v} \times \mathbf{B} \bigr) ,
\label{Lorentz} \\
{E_z}&= - \frac{S}{kw^2} [({x}+{y}) \sin{\phi_\parallel} + ({x}-{y})
\sin{(\phi_\parallel + \phi_p)}], \label{ez}\\[5pt]
{E_{x,y}}&= \frac{w_0 S}{2w}
[\,\cos{\phi_\perp} \pm \cos{(\phi_\perp + \phi_p)}],\\[5pt]
{B_z}&= \frac{S}{kw^2} [({x}-{y}) \sin{\phi_\parallel} - ({x}+{y})
\sin{(\phi_\parallel + \phi_p)}],\\[5pt]
B_x &= -E_y, \qquad B_y=E_x , \\[5pt]
S &= E_0 \exp{\left\{-\left( 2t/\tau \right)^4 \right\}}\ \exp{\left(-
\frac{{x}^2 + {y}^2}{{w}^2} \right)} , \label{sxy}
\end{align}
where $E_0$ is the field amplitude, $\phi_{\parallel}$ and $\phi_{\perp}$
are the phases of longitudinal and transversal field components:
\begin{eqnarray}
\phi_\parallel &=& \phi_\perp +\arctan{\left(\frac{z}{z_c}\right)},
\\
\phi_\perp &=& \phi_0 + \eta + \arctan{\left(\frac{z}{z_c}\right)} -
\frac{z}{z_c}\, \frac{x^2+y^2}{w^2},
\end{eqnarray}
with $\eta = \omega t -kz$ and $z_c = kw_0^2/2$ being the diffraction
length, and
\begin{eqnarray}\label{1}
    w^2 &=& w_0^2 [1+ (z/z_c)^2], %\\
%    \eta &=& \omega t -kz,
\end{eqnarray}
where $w_0$ is the minimum spot size of the laser beam at focus, $\phi_0$ is a
constant and $\phi_p$ is a parameter fixing the polarization of the field
($\phi_p=0$: linear, $\phi_p=\pi/2$: circular polarization).

\begin{figure}[t]
  \centering
\resizebox{0.45\textwidth}{!}{%
  \includegraphics{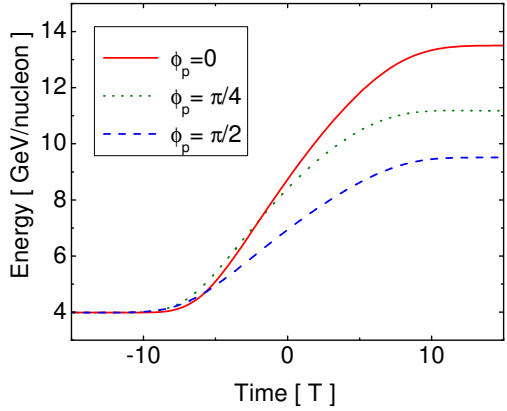}
  }
\caption{Time dependence of the energy of fully ionized gold ions Au$^{79+}$  for laser fields of different polarizations.}
   \label{fig:ac:time}
\end{figure}

\begin{figure}[t]
  \centering
  \resizebox{0.45\textwidth}{!}{%
  \includegraphics{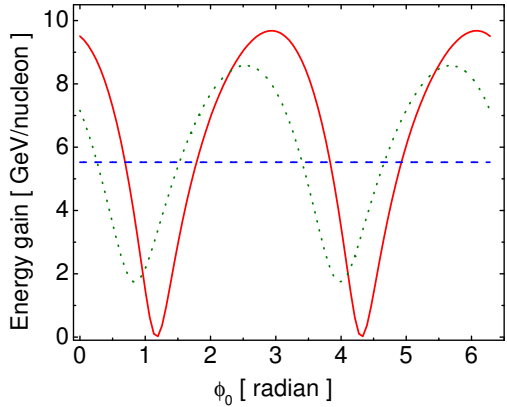}
  }
  \caption{Dependence of the ion energy gain $\Delta$ on the initial phase,
  $\phi_0$, for different polarizations: $\phi_p=0$ (solid line),
  $\phi_p=1$ (dashed line) and $\phi_p=0.5$ (dotted line).}
   \label{fig:ac:phi0}
\end{figure}

We suppose that at $t=t_0$ the ion is injected on beam axis at a distance $z_0$ from the focus with initial velocity $v_0$ along the axis. The Cauchy problem for Eq.\,(\ref{Lorentz}), $\mathbf{x}(t_0)= (0,0,z_0)$, $\mathbf{v}(t_0) = (0,0,v_0)$, is solved numerically using the standard Runge-Kutta method with basic parameters sets specified in Table~\ref{table2}.

\begin{table}[b]
\begin{center}
\caption{Basic set of parameters \label{table2}}
\begin{tabular}{|c|c|c|c|c|c|} \hline
I [W/cm$^2$]& $\lambda$ [nm]  & $\tau$ [T]  & $\phi_0$ & $\phi_p$ & $w_0$
[$\lambda$] \\
\hline  \rule{0em}{2.5ex}$5\times 10^{25}$ & 1000& 20  & 0& 0 & 100 \\
\hline \hline  \rule{0em}{2ex}e & M [GeV] & $v_0$ [c] &$t_0$ [$\tau$] & $z_0$ [$\lambda$] &\\
\hline \rule{0em}{2ex}79 & 179 & 0.968 & -1 & -100&\\ \hline
\end{tabular} 
\end{center}
\end{table}

The main characteristic of interest is the ion's energy gain, $\Delta$, during the period of field action.  The choice of ion type (fully ionized gold Au$^{79+}$) corresponds to conditions at the Dubna nuclotron \cite{Dubna} where the ion beam has final energy $4\,$GeV/nucleon.

\begin{figure}[t]
  \centering
  \resizebox{0.45\textwidth}{!}{%
  \includegraphics{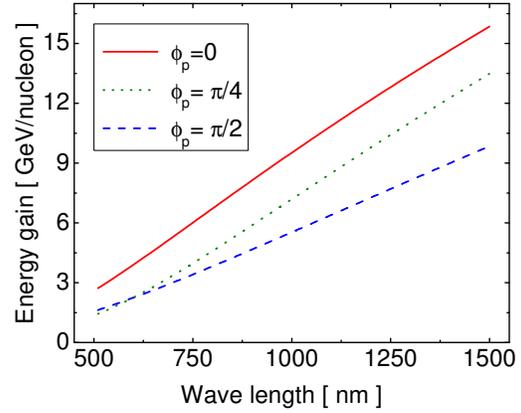}
  }
  \caption{The energy gain $\Delta$  vs. wavelength $\lambda$ of the laser 
field.  For the Phelix laser $\lambda= 1053$ nm \cite{Phelix}, while for the 
Astra laser $\lambda= 750 - 850$ nm \cite{Astra}.}   \label{fig:ac:1}
\end{figure}

\begin{figure}[t]
  \centering \resizebox{0.45\textwidth}{!}{%
  \includegraphics{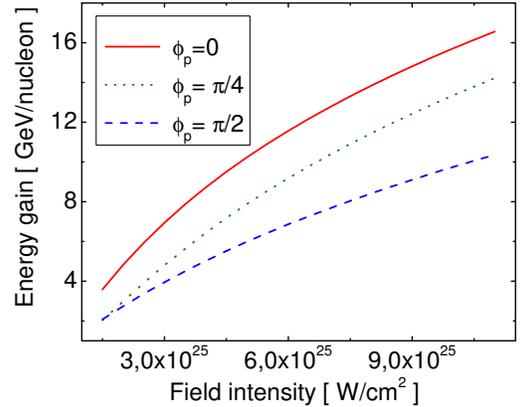}}
  \caption{The energy gain $\Delta$ vs. the intensity of laser field for 
different polarizations. The wavelength $\lambda = 1053$ nm corresponds to 
that of the Phelix laser \cite{Phelix}. }
   \label{fig:ac:3}
\end{figure}

\begin{figure}[t]
  \centering \resizebox{0.45\textwidth}{!}{%
  \includegraphics{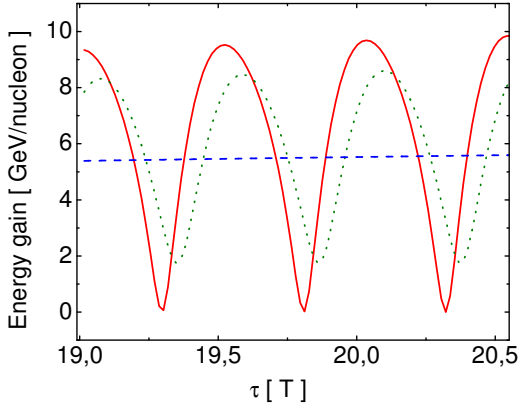}}
  \caption{The energy gain $\Delta$  vs.
the laser pulse duration at small time scale of order of the laser period.}
   \label{fig:ac:tau1}
\end{figure}

\begin{figure}[t]
  \centering \resizebox{0.45\textwidth}{!}{%
  \includegraphics{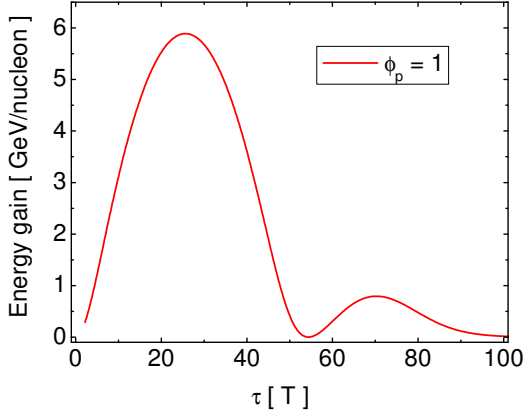}}
  \caption{Energy gain $\Delta$ vs. the laser pulse duration at large 
time scale; viz., much greater than the laser period.}
   \label{fig:ac:tau2}
\end{figure}

\begin{figure}[t]
  \centering \resizebox{0.45\textwidth}{!}{%
  \includegraphics{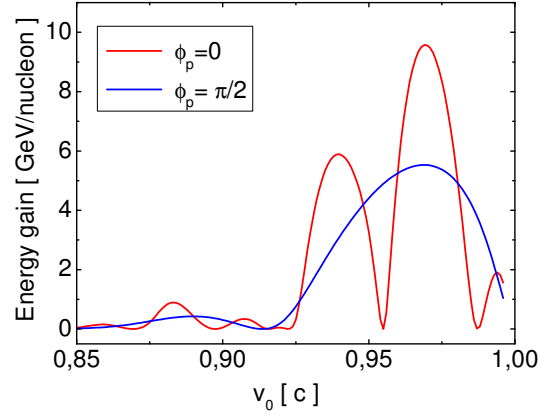}}
  \caption{Energy gain $\Delta$ vs. the initial velocity of the ion.}
   \label{fig:ac:v0}
\end{figure}

A picture typical of the ion energy's time dependence is depicted in Fig.\,\ref{fig:ac:time}.  The monotonic increase in energy indicates the heavy ion capture in the laser field.  In these conditions the relative phases have a
crucial influence on the acceleration mode.  Figure~\ref{fig:ac:phi0} illustrates the
dependence of the energy gain on the initial phase $\phi_0$.
The most significant influence of $\phi_0$ is observed for linear polarization and the least for circular polarization.  The permanent character of acceleration is also confirmed also by Fig.\,\ref{fig:ac:1} where the dependence of ion energy on laser wavelength is presented.  Larger wavelengths entail greater ion energy because of the increased interaction length.  The dependence of the energy gain on laser field intensity is similar, Fig.~\ref{fig:ac:3}.

It is interesting that the dependence of the energy gain on pulse duration, $\tau$, has an essentially non-monotonic character both at small time scales, of the order of the field period for linear polarization shown in Fig. \ref{fig:ac:tau1}, and at large time scales for circular polarization, shown in Fig. \ref{fig:ac:tau2}.  The curve in Fig.\,\ref{fig:ac:tau2} represents simultaneously the envelope for rapid energy oscillations with linear polarization.

The dependence of the energy on the initial ion velocity shows similar behavior, Fig.\,\ref{fig:ac:v0}.  The position of the local extrema on this curve can be changed by varying other parameters, such as the pulse duration or the field intensity.  Such a property can in principle be used for decreasing the velocity dispersion in the ion beam (``cooling''): maximum energy gain is enjoyed by ions within a narrow band of particular mean velocities, with little gain at the nodes.

\subsection{Two crossed beams scheme}
The use of special geometrical schemes to increase the laser acceleration efficiency has been discussed by many authors \cite{Haaland,AW,Salamin2003}.  The basic idea is to place the charged particles at the crossing point of two laser beams, which creates a longitudinally pulling electric field.  The resulting field is defined mainly by the  phase relations of the colliding laser beams.  To construct those relations can
be a difficult technical problem in the case of high-power systems \cite{AW}.  The two laser method is applied successfully in plasma accelerators, see also Ref. \cite{Faure}.

\begin{figure}[t]
  \centering \resizebox{0.45\textwidth}{!}{%
  \includegraphics{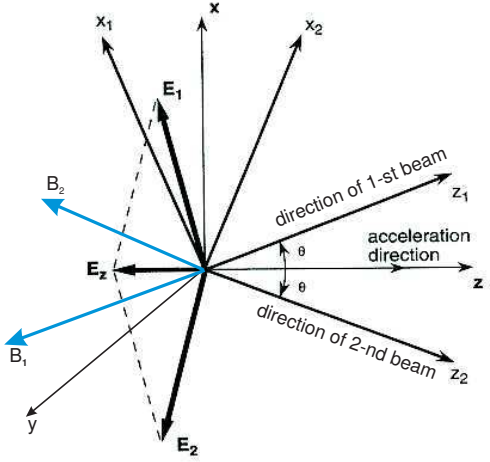}}
  \caption{Geometry of the crossed beam scheme according to Ref.\,\cite{Esarey93}
  in the linear polarization case.}
   \label{cross}
\end{figure}

Let us consider two laser beams propagated according to the model of Ref.\,\cite{Esarey93}, Fig.\,\ref{cross},
\begin{align}
E_x &= (E_{x1}+E_{x2}) \cos \theta + (E_{z1}-E_{z2}) \sin \theta , \nonumber \\
E_z &=  -(E_{x1} - E_{x2}) \sin \theta + (E_{z1} + E_{z2}) \cos \theta ,  \nonumber \\
x_{1,2}  &= x  \cos \theta  \mp z  \sin \theta ,  \nonumber \\
z_{1,2}  &=  \pm x  \sin \theta   + z  \cos \theta ,
\end{align}
where $\mathbf{E}_1(\mathbf{x_1},t)$ and $\mathbf{E}_2(\mathbf{x_2},t)$ are defined by Eqs. (\ref{ez})-(\ref{sxy}).  To obtain a purely axial field along the $z$ axis it is sufficient to set $E_{01} = - E_{02}$ \cite{Esarey93}.  This provides 1D-motion of the charge.

\begin{figure}[t]
  \centering \resizebox{0.45\textwidth}{!}{%
  \includegraphics{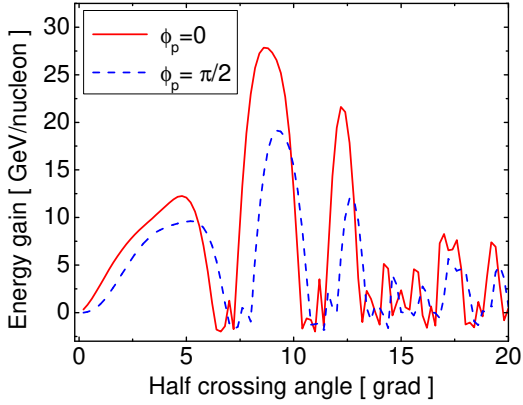}}
  \caption{Energy gain $\Delta$ vs. the half crossing angle.}
   \label{theta}
\end{figure}

The main characteristic of this scheme is the angular dependence of the energy gain, presented in Fig.\,\ref{theta} for a basic set of parameters.  This picture depends essentially on other parameters such as the initial position $z_0$ or the pulse duration $\tau$.  For example, the curve $\Delta (\theta)$ becomes smoother by decreasing $|z_0|$. In any case, an optimal value of crossing angle exists and is $\sim 20^\circ$, which correlates well with the result of Ref. \cite{Salamin2003}.  The efficiency of the crossed angle scheme is about three-times higher than the one beam scheme for the range of parameters considered.

\subsection{Background pair creation}

As we have seen, Sect.~\ref{sect:3:1}, a homogeneous electric field already creates a dense quasiparticle plasma at an intensity of the order of $10^{22}$ W/cm$^2$.  It is therefore necessary to account for the possible influence of this effect on the process of heavy ion acceleration.  Unfortunately, the direct calculation of pair creation in a field of Gaussian shape is at present beyond reach.  We provide an estimate via the field invariants
\begin{equation}\label{inv}
\mathbf{E}^2 - \mathbf{B}^2, \qquad \mathbf{E}\cdot\mathbf{B},
\end{equation}
supposing that the degree by which these quantities differ from zero indicates the probability for pair creation in the electromagnetic field.  This argument is based on the known property pair creation is absent for a plane wave.

\begin{figure}[t]
  \centering \resizebox{0.45\textwidth}{!}{%
  \includegraphics{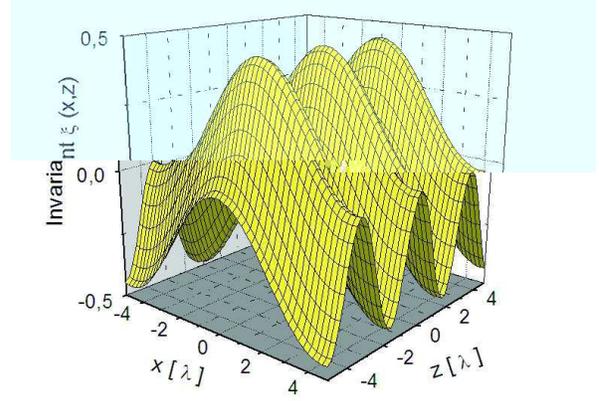}}
  \caption{The shape of the invariant $\xi(x,z)$ at $t=0$ for linearly
polarized crossed beams with $\theta=20^\circ$.
The beam ``waist'' $w_0$ is decreased here to $10\lambda$
  for presentation convenience.}
\label{inv1}
\end{figure}

In a sense a Gaussian beam is close to a plane wave because the longitudinal component is suppressed by a factor $\lambda/w_0 \ll 1$:
\begin{equation}\label{ksi}
\xi = \frac{\mathbf{E}^2 - \mathbf{B}^2}{{E}^2_{0}} \lesssim
\left(\frac{\lambda}{w_0}\right)^2.
\end{equation}
Hence it is natural to suppose that the creation rate is negligible for such a field.  In contrast to this, the crossed beams can produce a field of electric type locally by appropriate geometry and phasing; e.g., in the case of counter-propagating beams, when a standing wave is formed.  The arbitrary crossed angle scheme can only be estimated numerically, see Fig. \ref{inv1}.  We observe that regions with a sufficiently high value of $\xi$ occupy an appreciable part of field domain.  One can therefore expect the pair creation mechanism to work.  The presence of electron-positron pairs probably has a positive influence on the efficiency of ion acceleration.  However, this conjecture must be confirmed separately.

\section{Summary}
In this contribution we have described two types of nonperturbative QED effects for which the recent development of new, powerful laser technologies has opened up the possibility of experimental verification.
 
In the first part we reviewed an approach to the description of charged particle-antiparticle pair production in time-dependent external fields (time-dependent Schwinger effect) by a non-Markovian source term in a quantum kinetic equation, which is derived from a quantum-field theoretical formulation, as is summarized in the Appendix. 

As possible experimental signals of vacuum pair production we consider: $e^+e^-$ annihilation into $\gamma$-pairs; and refraction of a high-frequency probe laser beam by the $e^+e^-$ quasiparticle system, evolving in the focus of two counter-propagating
optical laser beams with intensities from 10$^{20}$-10$^{22}$ W/cm$^2$, achievable with present-day petawatt lasers, and up to the Schwinger limit, 10$^{29}$ W/cm$^2$,
to be reached at ELI.  For the highest intensities, the subthreshold production of  $\mu^+\mu^-$ and $\pi^+\pi^-$ pairs should manifest itself in observable
hard $\gamma$ photons from pair annihilation.

In the second part we summarized explorations of the question of whether high-intensity lasers can be used as ``boosters'' for ion beams in the few-GeV per nucleon range.  We provided estimates for the dependence of the energy gain on wavelength, geometry and intensity of the laser booster, which show that under optimal conditions a power of at least $10^{25}-10^{26}$ W/cm$^2$ is  required in order to achieve an energy gain in the range of a few GeV/nucleon. 

\section*{Acknowledgements}
D.B. thanks D. Habs and G. Mourou for enlightening discussions and encouragement during the ELI workshops, in particular the one at  Frauenw\"orth, October 2008. 
D.B. and G.R. acknowledge G. Gregori (Oxford) for his courageous step  to initiate an experiment at the Astra-Gemini laser at the Rutherford Appleton Laboratory, aiming at an experimental test of the nonstationary Schwinger  effect in the subcritical domain. 
D.B., A.V.P, G.R. and S.A.S. are grateful for support from the Helmholtz  Association for their participation at the Summer School on ``Dense Matter in  Heavy-Ion Collisions and Astrophysics'' in Dubna, July 14-26, 2008, where  this project has been started. 
This work was supported in part by the Department of Energy, Office of Nuclear 
Physics, contract no. DE-AC02-06CH11357 and by the Polish Ministry of Science 
and Higher Education under grant no.\ N N 202 0953 33.

\begin{appendix}

\section{Derivation of the kinetic equation}
\label{app:KE}

%%%%%%%%%%%%%%%%%%%%%%%%%%%%%%%%%%%%%%%%%%%%%%%%%%%%%%%%%%%%%%%%
In this section we demonstrate the derivation of a kinetic equation with a
method of the time-dependent canonical transformations \cite{Grib,Smol}
for particles obeying Fermi statistics.
The analogous derivation for the bosonic case is given in \cite{Smol}
We start from the QED Lagrangian
\begin{eqnarray}\label{2:2:1}
 {\cal L} = {\bar \psi} i\gamma^\mu(\partial_\mu+ieA_\mu)\psi
- m{\bar \psi}\psi-\frac{1}{4}F_{\mu\nu}F^{\mu\nu}\,,
\end{eqnarray}
where $F^{\mu\nu}$ is the field strength, the metric is taken as $g^{\mu\nu}
= {\rm diag}(1,-1,-1,-1)$ and for the $\gamma$-- matrices we use the
conventional definition \cite{bjorkendrell}. In the following we consider
the electromagnetic field as classical and quantize only the matter field.
Then the Dirac equation reads
\begin{eqnarray}\label{2:2:2}
(i\gamma^\mu\partial_\mu-e\gamma^\mu A_\mu-m)\psi(x)=0\,.
\end{eqnarray}
Using a simple field configuration with the vector potential in the
Hamiltonian gauge $A^\mu=(0,0,0,A(t))$ and a homogeneous electric field
$\mathbf{E}(t)=(0,0,E(t))$, $E(t)=-\dot A(t)$, one looks for the
solutions of the Eq.~(\ref{2:2:1}) in the form
\begin{multline}\label{2:2:10}
\psi^{(\pm)}_{\vp r}(x) =
\bigg[i\gamma^0\partial_0+\gamma^kp_k-e\gamma^3A(t)+m\bigg] \\ \times
\chi^{(\pm)}(\vp,t) \ R_r \ {\rm e}^{i\vp\bar x},
\end{multline}
where the superscript $(\pm)$ denotes eigenstates with the positive
and negative frequencies.  Herein the spinors $R_r$ $(r = 1,2)$ are
eigenvectors of the matrix $\gamma^0\gamma^3$ satisfying the
condition  $R^+_r R_s = 2\delta _{rs} \,.$ The functions
$\chi^{(\pm)}(\ve{q},t)$ obey the oscillator-type equation
\be \ds \lb{eq7}
{\ddot \chi}^{(\pm)}(\ve{q},t)+\big[\omega^2(\ve{q},t)+ie{\dot A}(t)\big] \
\chi^{(\pm)}(\ve{q},t)=0\,\,,
\ee
where we define  the total energy
$\var^2(\ve{q},t)=\varepsilon_{\trans}^2+p_{\along}^2(t)$,
the transverse energy $\varepsilon_{\trans}^2=m^2+\ve{q}^2_{\trans}$,
and the longitudinal momentum $p_{\along}(t)=q_{\along}-eA(t)$. The
solutions $\chi^{(\pm)}(\ve{q},t)$ of Eq.~(\ref{eq7}) for positive and
negative frequencies are fixed by their asymptotic behavior at
$t_0=t\rightarrow -\infty$, where $\dot A(t_0) = 0$. The field
operators $\psi(x)$ and ${\bar \psi}(x)$ can be decomposed by the
complete and orthonormalized set of spinor functions (\ref{2:2:10})
as follows:
\be \ds \lb{eq8}
\psi(x)=\sum\limits_{r,\ve{q}} \left[\psi^{(-)}_{\ve{q} r}(x)\ b_{\ve{q} r} +
\psi^{(+)}_{\ve{q} r}(x) \ d^+_{-\ve{q} r} \right] \,\,.
\ee
The operators $b_{\ve{q} r},b^+_{\ve{q} r}$ and $d_{\ve{q} r},d^+_{\ve{q} r}$
describe the annihilation and creation of particles and antiparticles
and obey the standard anticommutation rules. The time evolution
leads to the mixing of states with positive and negative energies
and, therefore, non-diagonal terms in the Hamiltonian corresponding
to Eq.~(\ref{2:2:2}) emerge. The diagonalization of the Hamiltonian,
which is equivalent to the transition to the quasiparticle representation,
is performed by the time-dependent Bogoliubov transformation
\bea \ds \lb{eq9}
\begin{array}{lcl}
b_{\ve{q} r}(t) &=& \alpha_\ve{q} (t)\ b_{\ve{q} r}
+ \beta_\ve{q} (t)\ d^+_{-\ve{q} r}\ , \\[7pt]
d_{\ve{q} r}(t) &=& \alpha_{-\ve{q}} (t) \
d_{\ve{q} r} - \beta_{-\ve{q}}(t) \ b^+_{-\ve{q} r}\ ,
\end{array}
\eea
with the imposed condition $|\alpha_\ve{q}(t)|^2+|\beta_\ve{q}(t)|^2=1\,.$
The new operators $b_{\ve{q} r}(t)$ and $d_{\ve{q} r}(t)$ describe the
processes of quasiparticle creation and annihilation. By the virtue of
Lagrange multipliers, one can find from the equations of motion
\re{eq7} that the coefficients in the Bogoliubov transformation
\re{eq9} are connected via the relations \cite{Grib}
\begin{equation} \ds \lb{eq10}
\begin{array}{lcl}  {\dot \alpha}_\ve{q}(t) &=& \quad
{\ds\frac{eE(t)\varepsilon_{\trans}}{2\var^2(\ve{q},t)}}\
\beta^*_\ve{q}(t) \ {\rm e}^{2i\theta(\ve{q},t_0,t)}\,\,, \\[15pt] {\dot
\beta}^*_\ve{q}(t)&=&-{\ds\frac{eE(t)\varepsilon_{\trans}}
{2\var^2(\ve{q},t)}}\
\alpha_\ve{q}(t) \ {\rm e}^{-2i\theta(\ve{q},t_0,t)}\,\,, \end{array}
\end{equation}
where the dynamical phase is defined as
\be \ds \lb{eq11}
\theta(\ve{q},t_0,t) = \int^t_{t_0}dt'\var(\ve{q},t')\,\,.
\ee
To absorb the dynamical phase it is convenient to introduce new
operators
\be \ds \lb{eq12}
\begin{array}{lcl}
B_{\ve{q} r}(t) &=& b_{\ve{q} r}(t) \  e^{-i\theta(\ve{q},t_0,t)}\,,\qquad \\
D_{\ve{q} r}(t) &=& d_{\ve{q} r}(t)\  e^ {-i\theta(\ve{q},t_0,t)}
\end{array}
\ee
which obey the anti-commutation relations:
\be \ds \lb{eq13}
\{B_{\ve{q} r}(t),B^+_{\ve{q} ' r'}(t)\}=\{D_{\ve{q} r}(t),D^+_{\ve{q} ' r'}(t)\}=
\delta_{rr'} \ \delta_{\ve{q} \ve{q} '} \, .
\ee
These operators satisfy the Heisenberg-type equations of motion
\bea \ds \lb{eq14}
\begin{array}{lcl} {\ds\frac{dB_{\ve{q} r}(t)}{dt}=-\frac{e
E(t)\varepsilon_{\trans}}{2\var^2(\ve{q},t)}} \ D^+_{-\ve{q} r}(t) +
i \ [H(t), \ B_{\ve{q} r}(t)]\,\,,
 \\ {\ds\frac{dD_{\ve{q} r}(t)}{dt}=\phantom{-}
\frac{e E(t)\varepsilon_{\trans}}
{2\var^2(\ve{q},t)}}\ B^+_{-\ve{q} r}(t)+i \ [H(t),\ D_{\ve{q} r}(t)] \ ,
\end{array}
\eea
where $H(t)$ is the Hamiltonian of the system of quasiparticles
\begin{multline}\lb{eq15}
H(t)= \sum_{r,\ve{q}} \var(\ve{q},t)\bigl[B^+_{\ve{q} r}(t) \ B_{\ve{q} r}(t) \\ -
D_{-\ve{q} r}(t) \ D^+_{-\ve{q} r}(t)\bigr]\,\, .
\end{multline}
The first term in the r.h.s. of Eqs.~(\ref{eq14}) arises because of
the transition from the representation \re{eq8} to the quasiparticle one
which are unitary non-equivalent.

Next we consider the evolution of the distribution function of
particles with the momentum $\ve{q}$ and spin $r$ defined as
\begin{multline} \lb{eq16}
f_r(\ve{q},t) =\, \langle 0_{in}|b^+_{\ve{q} r}(t) \ b_{\ve{q} r}(t)|0_{in}
\rangle\, \\  =\,
\langle 0_{in}|B^+_{\ve{q} r}(t) \ B_{\ve{q} r}(t)|0_{in} \rangle \ .
\end{multline}
According to the charge conservation the distribution functions for
particles and anti-particles are related as $f_r(\ve{q},t) = {\bar
f}_r(-\ve{q},t)$. Taking differentials in Eq.~(\ref{eq16}) with respect
to time $t$ we have
\be \ds \lb{eq17}
\frac{d f_r(\ve{q},t)}{dt}=
-\frac{eE(t)\,\varepsilon_{\trans}}{\var^2(\ve{q},t)} \ {\rm
Re}\{\Phi_r(\ve{q},t)\} \ .
\ee
Here the function $\Phi_r(\ve{q},t)=\langle 0_{in}|D_{-\ve{q} r}(t) \
B_{\ve{q} r}(t)|0_{in} \rangle$ describes the vacuum production of pairs
in the external electric field $E(t)$. Applying the equations of motion
(\ref{eq14}), one finds
\begin{multline} \ds\lb{eq18}
\frac{d\Phi_r(\ve{q},t)}{dt} = \frac{e E(t)\,\varepsilon_{\trans}}
{2\var^2(\ve{q},t)}\left[2f_r(\ve{q},t)-1\right] \\ -
2i\var(\ve{q},t) \ \Phi_r (\ve{q},t)\ .
\end{multline}
The solution of Eq.~(\ref{eq18}) with the initial condition
$\Phi_r(\ve{q},t_0)=0 $ may be written in the following integral form
\begin{multline}\ds\lb{eq19}
\Phi_r(\ve{q},t) = \frac{\varepsilon_{\trans}}{2}\int\limits_{t_0}^t dt'
\frac{e E(t')}{\var^2(\ve{q},t')}\left[2f_r(\ve{q},t')-1\right] \\ \times
\exp{[ 2i\theta(\ve{q},t',t)]}\,\,.
\end{multline}
Inserting this result into the r.h.s. of Eq.~(\ref{eq17}) we obtain
the anticipated kinetic equation
\begin{multline}\ds\lb{eq20}
\frac{df_r(\ve{q},t)}{dt} = \frac{e E(t)\varepsilon_{\trans}}
{2\var^2(\ve{q},t)}\int\limits_{t_0}^t \! dt' \,
\frac{e E(t')\varepsilon_{\trans}}{\var^2(\ve{q},t')} \\ \left[
1-2f_r(\ve{q},t')\right]\cos{ [2\,\theta(\ve{q},t',t)]}\ .
\end{multline}
Since the distribution function does not depend on spin, the
subscript $r$ can be dropped: $f_r \equiv f$.
The substitution $\ve{p} = \ve{q} - e \vec{A}(t)$, results in the KE
%(\ref{eq20}) to
\re{ke}.

The derived equation demonstrates several interesting
features, such as the dependence on particle longitudinal and
transverse momentum, the account for spin and statistics, and the
non-markovian character of the time evolution. The memory effects are
caused by the time integration over the statistical factor $(1-2f)$
and the non-local cosine function, while the structure of
the coefficient $\Delta (\ve{p})$ defines the shape of the momentum
distribution of created particles.

Based on microscopic dynamics, this kinetic equation is exact within the
approximation of a time-dependent homogeneous electric field and the neglect
of collisions. The source term is characterized by the following features:
\begin{itemize}
\item{The kinetic equations (\ref{ke}) are of non-Markovian type due
to the explicit dependence of the source terms on the whole pre-history via
the statistical factor $( 1 \pm 2f({\ve{p}},t) )$ for fermions or bosons,
respectively. The memory effect is expected to lead to a modification of
particle pair creation as compared to the (Markovian) low-density limit,
where the statistical factor is absent.}
\item{The presence of the dynamical phases,
$\theta(\vp ,t)$, under the cosine in the integrand (\ref{eq19})
 generates high frequency oscillations.}
\item{ The appearance of such a source term leads to entropy
production due to pair creation and therefore the
time reversal symmetry should be violated, but it does not result in any
monotonic entropy increase (in absence of collisions).}
\item{
The source term  and the distribution functions have a non-trivial momentum
dependence resulting in the fact that particles are produced not only at
rest as assumed in previous studies, e.g. Ref.\cite{Back}.}
\item{
In the low density limit and in the simple case of a constant electric field
we reproduce the pair production rate given by Schwinger's formula
\begin{multline}
\label{430} 
{S}^{{\,\rm cl}} 
= \lim_{t\to +\infty}(2\pi)^{-3} g \int d^3P \ {S}(\ve{p},t) \\ 
=\frac{e^2E^2}{4\pi^3}\exp \bigg(-\frac{\pi m^2}{|eE|}\bigg) \ .
\end{multline}}
\end{itemize}

In our approach the electric field is treated as a general time dependent 
field and hence there is no {\em a priori} limitation to constant fields. 
However our result allows to explore the influence of any time-dependent 
electric field on the pair creation process. 
It is important to note that in general this time dependence should be given 
by a selfconsistent solution of the coupled field equations, namely the Dirac 
(Klein-Gordon) equation and the Maxwell equation. 
This would incorporate back reactions as mentioned in the introduction.

Finally we remark that the source term is characterized by  two time scales:
the memory time
\begin{equation}
\tau_{mem}\sim \frac{\varepsilon_\perp}{eE}
\end{equation}
and the production interval
\begin{equation}
\tau_{prod}=1/<S>\,, 
\end{equation} 
with $<S>$ denoting the time averaged production rate. 
As long as $E \ll m^2/e < \varepsilon^2_\perp/e$,
the particle creation process is Markovian: $\tau_{mem} \ll \tau_{prod}$.

\end{appendix}

\begin{thebibliography}{99}

\bibitem{Sauter}
%\cite{Sauter:1931zz}
%\bibitem{Sauter:1931zz}
  F.~Sauter,
  %``Uber das Verhalten eines Elektrons im homogenen elektrischen Feld nach der
  %relativistischen Theorie Diracs,''
  Z.\ Phys.\  {\bf 69}, 742 (1931).
  %%CITATION = ZEPYA,69,742;%%

\bibitem{HE}
W.~Heisenberg and H.~Euler,
Z. Phys. \textbf{98}, 714 (1936).

\bibitem{JS}
%\cite{Schwinger:1951nm}
%\bibitem{Schwinger:1951nm}
  J.~S.~Schwinger,
  %``On gauge invariance and vacuum polarization,''
  Phys.\ Rev.\  {\bf 82}, 664 (1951).
  %%CITATION = PHRVA,82,664;%%

\bibitem{Grib}
A.~A.~Grib, S.~G.~Mamaev, and V.~M.~Mostepanenko,
\textit{Vacuum Quantum Effects in Strong External Fields}
(Friedmann Lab. Publ., St.-Petersburg, 1994)

\bibitem{Nick}
A.~I.~Nikishov,
Tr. Fiz. Inst. Akad. Nauk SSSR, \textbf{111}, 152 (1979).

\bibitem{Greiner}
W.~Greiner, B.~M\"uller, and J.~Rafelski,
\textit{Quantum Electrodynamics of Strong Fields}, (Springer, Berlin, 1985)
%\cite{Greiner:1985ce}
%\bibitem{Greiner:1985ce}
 % W.~Greiner, B.~Muller and J.~Rafelski,
  %``Quantum Electrodynamics Of Strong Fields,''
%\href{http://www.slac.stanford.edu/spires/find/hep/www?irn=1475096}{SPIRES entry}
%{\it  Berlin, Germany: Springer ( 1985) 594 P. ( Texts and Monographs In Physics)}

\bibitem{Fradkin}
E.~S.~Fradkin, D.~M.~Gitman, and S.~M.~Shvartsman,
\textit{Quantum Electrodynamics with Unstable Vacuum},
(Springer-Verlag, Berlin, 1991)

\bibitem{Smol}
%\cite{Schmidt:1998vi}
%\bibitem{Schmidt:1998vi}
  S.~M.~Schmidt, D.~Blaschke, G.~Ropke, S.~A.~Smolyansky, A.~V.~Prozorkevich
  and V.~D.~Toneev,
  %``A quantum kinetic equation for particle production in the Schwinger
  %mechanism,''
  Int.\ J.\ Mod.\ Phys.\  E {\bf 7}, 709 (1998).
%  [arXiv:hep-ph/9809227].
  %%CITATION = IMPAE,E7,709;%%

\bibitem{Rob}
%\cite{Roberts:2002py}
%\bibitem{Roberts:2002py}
  C.~D.~Roberts, S.~M.~Schmidt and D.~V.~Vinnik,
  %``Quantum effects with an X-ray free electron laser,''
  Phys.\ Rev.\ Lett.\  {\bf 89}, 153901 (2002).
%  [arXiv:nucl-th/0206004]
  %%CITATION = PRLTA,89,153901;%%

\bibitem{Pop} V.~S.~Popov, Phys. Lett. A \textbf{298}, 83 (2002)

\bibitem{Brezin70}
%\cite{Brezin:1970xf}
%\bibitem{Brezin:1970xf}
  E.~Brezin and C.~Itzykson,
  %``Pair production in vacuum by an alternating field,''
  Phys.\ Rev.\  D {\bf 2}, 1191 (1970).
  %%CITATION = PHRVA,D2,1191;%%E. Brezin and C. Itzykson,  Phys.~Rev. D


\bibitem{Casher}
A.~Casher, H.~Neuberger, and S.~Nussinov,
Phys. Rev. D \textbf{20},  179 (1979).
    %``Chromoelectric-flux-tube model of particle production"

\bibitem{magn}
V.~S.~Beskin, A.~V.~Gurevich, and Ya.~N.~Istomin,
\textit{Physics of the Pulsar Magnetosphere}
(Cambridge: Cambridge Univ. Press, 1993)

\bibitem{Ruffini}
%\cite{Ruffini:2003cr}
%\bibitem{Ruffini:2003cr}
  R.~Ruffini, L.~Vitagliano and S.~S.~Xue,
  %``On Plasma Oscillations in Strong Electric Fields,''
  Phys.\ Lett.\  B {\bf 559}, 12 (2003).
  %[arXiv:astro-ph/0302549]
  %%CITATION = PHLTA,B559,12;%%

\bibitem{focus}
B.~Richards and E.~Wolf,
Proc. Roy. Soc. A (London) \textbf{253}, 358 (1959).

\bibitem{Troup}
C.~J.~Troup and H.~S.~Perlman,
Phys. Rev. D \textbf{6}, 2299 (1972).

\bibitem{Marinov}
%\cite{Marinov:1977gq}
%\bibitem{Marinov:1977gq}
  M.~S.~Marinov and V.~S.~Popov,
  %``Electron-Positron Pair Creation From Vacuum Induced By Variable Electric
  %Field,''
  Fortsch.\ Phys.\  {\bf 25}, 373 (1977).
  %%CITATION = FPYKA,25,373;%%

\bibitem{Bunkin}
F.~V.~Bunkin and I.~I.~Tugov,
Dokl. Akad. Nauk. SSSR \textbf{187}, 541 (1964);
[Sov. Phys. Dokl. \textbf{14}, 678 (1969)].

\bibitem{BulanovSS}
S.~S.~Bulanov, Phys. Rev. E \textbf{69}, 0326408 (2004);\\
%
%\cite{Bulanov:2004de}
%\bibitem{Bulanov:2004de}
S.~S.~Bulanov, N.~B.~Narozhny, V.~D.~Mur and V.~S.~Popov,
  %``On e+ e- pair production by a focused laser pulse in vacuum,''
  Phys.\ Lett.\  A {\bf 330}, 1 (2004).
%  [arXiv:hep-ph/0403163];
  %%CITATION = PHLTA,A330,1;%%
%
%\cite{Bulanov:2004kh}
%\bibitem{Bulanov:2004kh}
  S.~S.~Bulanov, A.~M.~Fedotov and F.~Pegoraro,
  %``Damping of electromagnetic waves due to electron positron pair
  %production,''
  Phys.\ Rev.\  E {\bf 71}, 016404 (2005).
%  [arXiv:hep-ph/0409301]
  %%CITATION = PHRVA,E71,016404;%%


\bibitem{Mou}
G.~A.~Mourou, C.~P.~J.~Barty, and M.~D.~Perry,
Phys. Today \textbf{51}, 22 (1998).

\bibitem{BulanovSV}
S.~V.~Bulanov, T.~Esirkepov, and T.~Tajima,
Phys. Rev. Lett.  \textbf{91}, 085001 (2003);
Erratum \textbf{92}, 159901 (2004).

\bibitem{ELI}
http://www.extreme-light-infrastructure.eu/

\bibitem{Ring}
%\cite{Ringwald:2001ib}
%\bibitem{Ringwald:2001ib}
  A.~Ringwald,
  %``Pair production from vacuum at the focus of an X-ray free electron
  %laser,''
  Phys.\ Lett.\  B {\bf 510}, 107 (2001).
%  [arXiv:hep-ph/0103185]
  %%CITATION = PHLTA,B510,107;%%

\bibitem{XFEL}
http://xfel.desy.de/

\bibitem{PRL06}
%\cite{Blaschke:2005hs}
%\bibitem{Blaschke:2005hs}
  D.~B.~Blaschke, A.~V.~Prozorkevich, C.~D.~Roberts, S.~M.~Schmidt and
S.~A.~Smolyansky,
  %``Pair production and optical lasers,''
  Phys.\ Rev.\ Lett.\  {\bf 96}, 140402 (2006).
%  [arXiv:nucl-th/0511085]
  %%CITATION = PRLTA,96,140402;%%



\bibitem{OR}
V.~N.~Pervushin, V.~V.~Skokov, A.~V.~Reichel, S.~A.~Smolyansky,
A.~V.~Prozorkevich,
 %"The kinetic description of vacuum particle creation in the oscillator representation"
 Int. J. Mod. Phys. A \textbf{20}, 5689 (2005).
 % [arXiv:hep-th/0307200]

\bibitem{our}
S.~Schmidt, D.~Blaschke, G.~R\"opke,
A.~V.~Prozorkevich, S.~A.~Smolyansky, and V.~D.~Toneev,
Phys. Rev. D \textbf{59}, 094005 (1999).

\bibitem{Bloch:1999eu}
  J.~C.~R.~Bloch, V.~A.~Mizerny, A.~V.~Prozorkevich, C.~D.~Roberts, S.~M.~Schmidt, S.~A.~Smolyansky and D.~V.~Vinnik,
  %``Pair creation: Back-reactions and damping,''
  Phys.\ Rev.\  D {\bf 60}, 116011 (1999).
%  [arXiv:nucl-th/9907027]
  %%CITATION = PHRVA,D60,116011;%%


\bibitem{quark}
%\cite{Prozorkevich:2004yp}
%\bibitem{Prozorkevich:2004yp}
  A.~V.~Prozorkevich, S.~A.~Smolyansky, V.~V.~Skokov and E.~E.~Zabrodin,
  %``Vacuum creation of quarks at the time scale of QGP thermalization and
  %strangeness enhancement in heavy-ion collisions,''
  Phys.\ Lett.\  B {\bf 583}, 103 (2004).
 % [arXiv:nucl-th/0401056]
  %%CITATION = PHLTA,B583,103;%%


\bibitem{bastirev}
%\cite{Roberts:2000aa}
%\bibitem{Roberts:2000aa}
  C.~D.~Roberts and S.~M.~Schmidt,
  %``Dyson-Schwinger equations: Density, temperature and continuum strong
  %QCD,''
  Prog.\ Part.\ Nucl.\ Phys.\  {\bf 45}, S1 (2000).
%  [arXiv:nucl-th/0005064]
  %%CITATION = PPNPD,45,S1;%%



\bibitem{Mamaev78}
%\cite{Mamaev:1978hf}
%\bibitem{Mamaev:1978hf}
  S.~G.~Mamaev and V.~M.~Mostepanenko,
  %``Renormalization Of Gravitational Constant And Creation Of Fermions By
  %Nonstationary Gravitational Field,''
  Yad.\ Fiz.\  {\bf 28}, 1640 (1978).
  %%CITATION = YAFIA,28,1640;%%

\bibitem{Zeld71}
%\cite{Zeldovich:1971mw}
%\bibitem{Zeldovich:1971mw}
  Y.~B.~Zeldovich and A.~A.~Starobinsky,
%``Particle production and vacuum polarization in an anisotropic gravitational
  %field,''
  Sov.\ Phys.\ JETP {\bf 34}, 1159 (1972);
  [Zh.\ Eksp.\ Teor.\ Fiz.\  {\bf 61}, 2161 (1971)].
  %%CITATION = ZETFA,61,2161;%%



\bibitem{Jena} B. Liesfeld,
J.~Bernhardt, K.-U.~Amthor, H.~Schwoerer, and R.~Sauerbrey,
Appl. Phys. Lett. {\bf 86}, 161107 (2005).

\bibitem{SLAC}
 %\cite{Bula:1996st}
 %\bibitem{Bula:1996st}
  C.~Bula {\it et al.}  [E144 Collaboration],
  %``Observation of nonlinear effects in Compton scattering,''
  Phys.\ Rev.\ Lett.\  {\bf 76}, 3116 (1996);\\
  %%CITATION = PRLTA,76,3116;%%
 %\cite{Burke:1997ew}
%\bibitem{Burke:1997ew}
  D.~L.~Burke {\it et al.},
  %``Positron production in multiphoton light-by-light scattering,''
  Phys.\ Rev.\ Lett.\  {\bf 79}, 1626 (1997).
  %%CITATION = PRLTA,79,1626;%%

\bibitem{Sh} 
J.~W.~Shearer, J.~Garrison, J.~Wong, and J.~E.~Swain, 
Phys. Rev. A \textbf{8}, 1582 (1973).

\bibitem{Av02}
%\cite{Avetissian:2002zz}
%\bibitem{Avetissian:2002zz}
  H.~K.~Avetissian and G.~F.~Mkrtchian,
  %``Coherent x-ray source due to the quantum reflection of an electron beam
  %from a laser-field phase lattice,''
  Phys.\ Rev.\  E {\bf 65}, 016506 (2002).
  %%CITATION = PHRVA,E65,016506;%%

\bibitem{Esarey93}
%\cite{Esarey:1993zz}
%\bibitem{Esarey:1993zz}
  E.~Esarey, S.~K.~Ride and P.~Sprangle,
  %``Nonlinear Thomson scattering of intense laser pulses from beams and
  %plasmas,''
  Phys.\ Rev.\  E {\bf 48}, 3003 (1993).
  %%CITATION = PHRVA,E48,3003;%%


\bibitem{Landau4} 
L.~D.~Landau and E.~M.~Lifshitz, 
\textit{Quantum Electrodynamics}, 
Volume 4 (Course of Theoretical Physics Series),
Butterworth-Heinemann; 2 ed. (1982)

\bibitem{Ivanov}
%\cite{Ivanov:2005ay}
%\bibitem{Ivanov:2005ay}
  D.~Y.~Ivanov, G.~L.~Kotkin and V.~G.~Serbo,
  %``Polarization effects in the non-linear Compton scattering,''
  arXiv:hep-ph/0501263
  %%CITATION = HEP-PH/0501263;%%

\bibitem{Heinzl_06}
%\cite{Heinzl:2006xc}
%\bibitem{Heinzl:2006xc}
  T.~Heinzl, B.~Liesfeld, K.~U.~Amthor, H.~Schwoerer, R.~Sauerbrey and A.~Wipf,
  %``On the observation of vacuum birefringence,''
  Opt.\ Commun.\  {\bf 267}, 318 (2006).
%  [arXiv:hep-ph/0601076]
  %%CITATION = OPCOB,267,318;%%

\bibitem{Astra} Astra Gemini Project,
http://www.clf.rl.ac.uk / Facilities / AstraWeb / AstraGeminiHome.htm

\bibitem{Heinzl_08}
%\cite{Heinzl:2008wh}
%\bibitem{Heinzl:2008wh}
  T.~Heinzl and A.~Ilderton,
  %``Extreme field physics and QED,''
  arXiv:0809.3348 [hep-ph];
  %%CITATION = ARXIV:0809.3348;%%
%\cite{Heinzl:2008an}
%\bibitem{Heinzl:2008an}
  T.~Heinzl and A.~Ilderton,
  %``Exploring high-intensity QED at ELI,''
  arXiv:0811.1960 [hep-ph]
  %%CITATION = ARXIV:0811.1960;%%

\bibitem{Piazza06}
%\cite{DiPiazza:2006pr}
%\bibitem{DiPiazza:2006pr}
  A.~Di Piazza, K.~Z.~Hatsagortsyan and C.~H.~Keitel,
  %``Light diffraction by a strong standing electromagnetic wave,''
  Phys.\ Rev.\ Lett.\  {\bf 97}, 083603 (2006).
  [arXiv:hep-ph/0602039]
  %%CITATION = PRLTA,97,083603;%%

\bibitem{KleinNigam}
%\cite{Klein:1964zz}
%\bibitem{Klein:1964zz}
  J.~J.~Klein and B.~P.~Nigam,
  %``Dichroism of the Vacuum,''
  Phys.\ Rev.\  {\bf 136}, B1540 (1964).
  %%CITATION = PHRVA,136,B1540;%%

\bibitem{Baier67} 
R.~Baier and P.~Breitenlohner, 
Acta Phys. Austr. \textbf{25}, 212 (1967).

\bibitem{Brezin71}
%\cite{Brezin:1971nd}
%\bibitem{Brezin:1971nd}
  E.~Brezin and C.~Itzykson,
  %``Polarization phenomena in vacuum nonlinear electrodynamics,''
  Phys.\ Rev.\  D {\bf 3}, 618 (1971).
  %%CITATION = PHRVA,D3,618;%%

\bibitem{Birula70}
%\cite{BialynickaBirula:1970vy}
%\bibitem{BialynickaBirula:1970vy}
  Z.~Bialynicka-Birula and I.~Bialynicki-Birula,
 %``Nonlinear effects in Quantum Electrodynamics. Photon propagation and photon
  %splitting in an external field,''
  Phys.\ Rev.\  D {\bf 2}, 2341 (1970).
  %%CITATION = PHRVA,D2,2341;%%

\bibitem{Landau8} L.~D.~Landau and E.~M.~Lifshitz,
\textit{Electrodynamics of Continious Media},
Volume 8 (Course of Theoretical Physics Series), Pergamon Press, 1984.

\bibitem{Landau10} L.~D.~Landau and E.~M.~Lifshitz, \textit{Physical Kinetics},
Volume 10 (Course of Theoretical Physics Series), Butterworth-Heinemann, 1981.

\bibitem{Filatov} 
A.~V.~Filatov, A.~V.~Prozorkevich and S.~A.~Smolyansky, in
Proceedings of SPIE, \textbf{6165}, 
Eds. Vladimir L.~Derbov, Leonid A.~Melnikov, Lev M.~Babkov, (2006) p. 616509

\bibitem{Skokov}
%\cite{Pervushin:2006vh}
%\bibitem{Pervushin:2006vh}
  V.~N.~Pervushin and V.~V.~Skokov,
  %``Kinetic description of fermion production in the oscillator
  %representation,''
  Acta Phys.\ Polon.\  B {\bf 37}, 2587 (2006).
  [arXiv:astro-ph/0611780]
  %%CITATION = APPOA,B37,2587;%%


\bibitem{LCLS} 
J.~Arthur et al,  [LCLS Design Study Group Collaboration], 
``Linac coherent light source (LCLS) design study report,''
SLAC-R-0521 (1998)


\bibitem{desy} TESLA Technical Design Report, Part V: The X-Ray Free Electron
Laser, edited by G.~Ma\-ter\-lik and Th.\ Tschentscher, available at
http://tesla.desy.de/new\_pages/TDR\_CD/start.html

\bibitem{Chen} P.~Chen and C.~Pellegrini, in \textit{Proceedings of the 15th
Advanced ICFA Beam Dynamics Workshop on Quantum Aspects of Beam Physics,
Monterey, CA, 1998}, edited by P.~Chen (World Scientific, Singapore, 1999)
p.~571


\bibitem{Fried}
%\cite{Fried:2001ga}
%\bibitem{Fried:2001ga}
  H.~M.~Fried, Y.~Gabellini, B.~H.~J.~McKellar and J.~Avan,
  %``Pair production via crossed lasers,''
  Phys.\ Rev.\  D {\bf 63}, 125001 (2001).
  %%CITATION = PHRVA,D63,125001;%%

\bibitem{Popov01}
%\cite{Popov:2001ak}
%\bibitem{Popov:2001ak}
  V.~S.~Popov,
  %``On Schwinger mechanism of e+ e- pair production from vacuum by the  field
  %of optical and X-ray lasers,''
  JETP Lett.\  {\bf 74}, 133 (2001);
  [Pisma Zh.\ Eksp.\ Teor.\ Fiz.\  {\bf 74}, 151 (2001)].
  %%CITATION = ZFPRA,74,151;%%

\bibitem{bastiprl}
%\cite{Alkofer:2001ik}
%\bibitem{Alkofer:2001ik}
  R.~Alkofer, M.~B.~Hecht, C.~D.~Roberts, S.~M.~Schmidt and D.~V.~Vinnik,
  %``Pair Creation and an X-ray Free Electron Laser,''
  Phys.\ Rev.\ Lett.\  {\bf 87}, 193902 (2001).
%  [arXiv:nucl-th/0108046].
  %%CITATION = PRLTA,87,193902;%%


\bibitem{Popov72}
%\cite{Popov:1973az}
%\bibitem{Popov:1973az}
  V.~S.~Popov and M.~S.~Marinov,
  %``E+ e- pair production in variable electric field,''
  Yad.\ Fiz.\  {\bf 16}, 809 (1972).
  %%CITATION = YAFIA,16,809;%%

\bibitem{NN73} 
N.~B.~Narozhnyi and A.~I.~Nikishov,
%``Pair Production by a Periodic Electric Field,''
Zh.\ Eksp.\ Teor.\ Fiz.\ {\bf 65}, 862 (1973) [Sov.\ Phys.\ JETP {\bf 38},
427 (1974)].


\bibitem{kme}
%\cite{Kluger:1998bm}
%\bibitem{Kluger:1998bm}
  Y.~Kluger, E.~Mottola and J.~M.~Eisenberg,
  %``The quantum Vlasov equation and its Markov limit,''
  Phys.\ Rev.\  D {\bf 58}, 125015 (1998).
%  [arXiv:hep-ph/9803372].
  %%CITATION = PHRVA,D58,125015;%%


\bibitem{Paris}
%\cite{Mourou:2006zz}
%\bibitem{Mourou:2006zz}
  G.~A.~Mourou, T.~Tajima and S.~V.~Bulanov,
  %``Optics in the relativistic regime,''
  Rev.\ Mod.\ Phys.\  {\bf 78}, 309 (2006).
  %%CITATION = RMPHA,78,309;%%

\bibitem{AB} 
A.~I.~Akhiezer and V.~B.~Berestezky, 
\textit{Quantum Electrodynamics}, (Moscow, Nauka, 1951)

\bibitem{Piazza07} 
A.~Di~ Piazza, K.~Z.~Hatsagortsyan, C.~H.~Keitel,
Physics of Plasmas, \textbf{14}, 032102 (2007).


\bibitem{Nik64} 
A.~I.~Nikishov and V.~I.~Ritus,
Sov. Phys. JETP {\bf 19}, 529 (1964).

\bibitem{Keitel_mu}
%\cite{Muller:2006fj}
%\bibitem{Muller:2006fj}
  C.~M\"uller, K.~Z.~Hatsagortsyan and C.~H.~Keitel,
  %``Muon pair creation from positronium in a circularly polarized laser
  %field,''
  Phys.\ Rev.\  D {\bf 74}, 074017 (2006).
%  [arXiv:physics/0602106];
  %%CITATION = PHRVA,D74,074017;%%
%\cite{Hatsagortsyan:2006ww}
%\bibitem{Hatsagortsyan:2006ww}
  K.~Z.~Hatsagortsyan, C.~M\"uller and C.~H.~Keitel,
  %``Microscopic laser-driven high-energy colliders,''
  Europhys.\ Lett.\  {\bf 76}, 29 (2006).
%  [arXiv:physics/0602093]
  %%CITATION = EULEE,76,29;%%


\bibitem{ILIAS07} 
D.~B.~Blaschke, A.~V.~Prozorkevich, S.~A.~Smolyansky,
D.~S.~Shkirmanov, M.~Chubaryan, in
``ILIAS, Ion and Laser beam Interaction and Application Studies,
Progress Report No 2 of the PHELIX theory group'', Eds. P. Mulser
and T. Schlegel, (GSI Report 2007-03), p.~34

\bibitem{ILIAS08} 
D.~B.~Blaschke, A.~V.~Filatov, A.~V.~Prozorkevich, and D.~S.~Shkirmanov, 
%Heavy ions in a high-power laser beam, 
in ``ILIAS, Ion and Laser beam Interaction and Application Studies, 
Progress Report No 3 of the PHELIX theory group'', 
Ed. P. Mulser, (GSI Report 2008-05), p.~54

\bibitem{Shimoda} 
K.~Shimoda, Appl. Opt. \textbf{1}, 33 (1962).

\bibitem{Ashkin}
%\cite{Ashkin:1970mb}
%\bibitem{Ashkin:1970mb}
  A.~Ashkin,
  %``Acceleration and trapping of particles by radiation pressure,''
  Phys.\ Rev.\ Lett.\  {\bf 24}, 156 (1970);
  %%CITATION = PRLTA,24,156;%%;
   ibid, {\bf 25},  1321 (1970).

\bibitem{Tajima}
%\cite{Tajima:1979bn}
%\bibitem{Tajima:1979bn}
  T.~Tajima and J.~M.~Dawson,
  %``Laser electron accelerator,''
  Phys.\ Rev.\ Lett.\  {\bf 43}, 267 (1979).
  %%CITATION = PRLTA,43,267;%%

\bibitem{Malka}
%\cite{Malka:2002nf}
%\bibitem{Malka:2002nf}
  V.~Malka {\it et al.},
 %``Electron acceleration by a wake field forced by an intense ultrashort laser
  %pulse,''
  Science {\bf 298}, 1596 (2002).
  %%CITATION = SCIEA,298,1596;%%

\bibitem{Patel}
%\cite{Patel:2007zza}
%\bibitem{Patel:2007zza}
  N.~Patel,
  %``Accelerator physics: The plasma revolution,''
  Nature {\bf 449N7159}, 133 (2007).
  %%CITATION = NATUA,449N7159,133;%%

\bibitem{target} 
S.~Atzeni, M.~Temporal, and J.~ J.~ Honrubia, 
Nucl. Fusion \textbf{42}, L1 (2002).

\bibitem{Ren} 
J.~Ren, W.~Cheng, S.~Li and S.~Suckewer, 
Nature Physics \textbf{3}, 732 (2007).

\bibitem{Dubna} http://nucloserv.jinr.ru/index.htm

\bibitem{BT} 
A.~Bahari, V.~D.~Taranukhin, Quantum Electronics {\bf 34}, 129 (2004).

\bibitem{Blumen} 
I.~Blumenfeld et al. Nature \textbf{445}, 741 (2007).

\bibitem{Salam2006}
%\cite{Salamin:2006ff}
%\bibitem{Salamin:2006ff}
  Y.~I.~Salamin, S.~X.~Hu, K.~Z.~Hatsagortsyan and C.~H.~Keitel,
  %``Relativistic high-power laser-matter interactions,''
  Phys.\ Rept.\  {\bf 427}, 41 (2006).
  %%CITATION = PRPLC,427,41;%%

\bibitem{Phelix} http://www.gsi.de/forschung/phelix/

\bibitem{Haaland}
C.~M.~Haaland, Opt. Comm. {\bf 114}, 280 (1995).

\bibitem{AW} 
A.~Aiello and  H.~Woerdman, arxiv:0710.1643

\bibitem{Salamin2003}
%\cite{Salamin:2003gc}
%\bibitem{Salamin:2003gc}
  Y.~I.~Salamin, G.~R.~Mocken and C.~H.~Keitel,
 %``Relativistic electron dynamics in intense crossed laser beams: Acceleration
  %and Compton harmonics,''
  Phys.\ Rev.\  E {\bf 67}, 016501 (2003).
  %%CITATION = PHRVA,E67,016501;%%

\bibitem{Faure}
%\cite{Faure:2006js}
%\bibitem{Faure:2006js}
  J.~Faure, C.~Rechatin, A.~Norlin, A.~Lifschitz, Y.~Glinec and V.~Malka,
  %``Controlled injection and acceleration of electrons in plasma wakefields by
  %colliding laser pulses,''
  Nature {\bf 444}, 737 (2006).
  %%CITATION = NATUA,444,737;%%

\bibitem{bjorkendrell}
J.~D.~Bjorken and S.~D.~Drell,
{\em Relativistic Quantum Mechanics} (McGraw-Hill, New York, 1964)

\bibitem{Back}
%\cite{Kluger:1992gb}
%\bibitem{Kluger:1992gb}
  Y.~Kluger, J.~M.~Eisenberg, B.~Svetitsky, F.~Cooper and E.~Mottola,
  %``Fermion Pair Production In A Strong Electric Field,''
  Phys.\ Rev.\  D {\bf 45}, 4659 (1992).
  %%CITATION = PHRVA,D45,4659;%%

\end{thebibliography}
\end{document}